\documentclass{iopjournal}

\usepackage{iftex}
\ifPDFTeX
  \usepackage[utf8]{inputenc}
\fi
\usepackage[T1]{fontenc}
\usepackage{amsmath,amssymb}
\newcommand{\journalheadingfont}{\fontfamily{qtm}\selectfont}
\newcommand{\affiliationmark}[1]{\raisebox{0.65ex}{\fontsize{8}{8}\selectfont #1}}
\let\originalmathbb\mathbb
\renewcommand{\mathbb}[1]{%
  \ifx#1e\mathrm{e}\else
    \ifx#1i\mathrm{i}\else\originalmathbb{#1}\fi
  \fi}
\usepackage{longtable,booktabs,array,multirow,calc}
\usepackage{graphicx}
\usepackage{caption}
\usepackage{xurl}
\usepackage[a4paper,left=16.8mm,right=15.5mm,top=23.5mm,bottom=23.5mm,headsep=5mm,footskip=8mm]{geometry}
\usepackage{ragged2e}
\usepackage{dblfloatfix}
\usepackage{placeins}
\usepackage{enumitem}
\usepackage{balance}

\makeatletter
\renewcommand{\title}[1]{{\exhyphenpenalty=10000\hyphenpenalty=10000
  \journalheadingfont\fontsize{24}{29}\selectfont\noindent\raggedright
  #1\par}\suppressfloats[t]}
\renewcommand\section{\@startsection {section}{1}{\z@}%
  {-7pt}{7pt}%
  {\reset@font\journalheadingfont\normalsize\bfseries\upshape\raggedright}}
\renewcommand\subsection{\@startsection{subsection}{2}{\z@}%
  {-7pt}{7pt}%
  {\reset@font\journalheadingfont\small\bfseries\upshape\raggedright}}
\renewcommand\subsubsection{\@startsection{subsubsection}{3}{\z@}%
  {-6pt}{6pt}%
  {\reset@font\journalheadingfont\small\bfseries\upshape}}
\renewenvironment{abstract}{%
      \vspace{12pt}
      \noindent \fontsize{9}{11.2}\selectfont {\bfseries Abstract}\\
      \rm\ignorespaces \raggedright}{\vspace{2mm}}
\makeatother

\setlist[enumerate]{leftmargin=*,labelsep=0.45em,itemsep=0.35ex,topsep=0pt,parsep=0pt}
\begin{document}
\pagestyle{plain}

\twocolumn[
\begin{@twocolumnfalse}
\title{Anisotropic Maxwell neural operator for rapid parametric full-wave modelling of ion cyclotron resonance heating}

\vspace{18pt}
{\journalheadingfont\fontsize{11}{14}\selectfont\bfseries\raggedright
Heng Zhang\affiliationmark{1,2,3}, Xu Wang\affiliationmark{1}, Jiayi Li\affiliationmark{1,2}, Miao Zhang\affiliationmark{4}, Jiahui Zhang\affiliationmark{4}, Kaihao Wang\affiliationmark{1}, Yangdi Yi\affiliationmark{1}, Qin Hang\affiliationmark{1,2,3,*}, and
Xinjun Zhang\affiliationmark{5}\par}

\vspace{8pt}
{\fontsize{10}{12}\selectfont\raggedright
\affiliationmark{1} School of Computer Science and Technology, Chongqing University of Posts and Telecommunications, Chongqing 400065, China\\
\affiliationmark{2} Center for Scientific Intelligence Innovation, University of Science and Technology of China, Hefei 230026, China\\
\affiliationmark{3} Institute of Advanced Technology, University of Science and Technology of China, Hefei 230088, China\\
\affiliationmark{4} School of Aerospace Science and Technology, Xidian University, Xi’an 710071, China\\
\affiliationmark{5} Institute of Plasma Physics, Hefei Institutes of Physical Science, Chinese Academy of Sciences, Hefei 230031, China\par}

\vspace{12pt}
{\fontsize{10}{12}\selectfont\raggedright E-mail: hangqin@cqupt.edu.cn\par}

\begin{abstract}
Full-wave calculations of ion cyclotron resonance heating (ICRH) under different plasma dielectric conditions generally require repeated assembly and solution of large-scale discretised systems for individual cases, limiting parameter sweeps and multi-case response analysis. We therefore propose an anisotropic Maxwell neural operator (AMNO) for rapid parametric modelling of ICRH full-wave responses for the Experimental Advanced Superconducting Tokamak (EAST), which learns, within the one-parameter dielectric-field family generated by varying the hydrogen minority fraction \(X_{H}\) over 0.01--0.05 under otherwise fixed physical and computational settings, a shared solution operator from the corresponding spatially varying complex anisotropic dielectric-tensor field to the three-component complex electric field under frequency-domain Maxwell constraints. It jointly represents global spatial coupling through spectral operator layers and local fine-scale responses, and combines sparse reference-field supervision with the frequency-domain Maxwell-equation residual in learning a function-to-function solution operator shared across dielectric conditions. Systematic comparisons with COMSOL reference solutions of the same frequency-domain Maxwell-dielectric model for the EAST configuration show that AMNO reconstructs the principal spatial and spectral features across dielectric conditions and maintains stable accuracy for unseen interpolation test cases. With reference-field points reduced to 7.5\% of the dense full-wave set, AMNO reduces the relative \(L_{2}\) errors by 66.1\%--89.9\% compared with a sparsely supervised Fourier neural operator (FNO-Sparse) under the same sparse reference-field supervision and requires only about 0.25 s for single-case inference. These results show that AMNO substantially reduces dependence on dense reference-field supervision while enabling the transition from case-by-case full-wave solutions to subsecond parametric complex-field inference, providing a physics-constrained and data-efficient surrogate for rapid in-range \(X_{H}\) sweeps and cross-case response analysis within the modelled EAST configuration.
\end{abstract}

\vspace{2mm}
{\color{gray}\hrule}
\vspace{4mm}
\end{@twocolumnfalse}
]
\fontsize{9}{11.2}\selectfont
\justifying

\begin{figure*}[!b]
\centering
\includegraphics[width=1.00\textwidth,height=0.68\textheight,keepaspectratio]{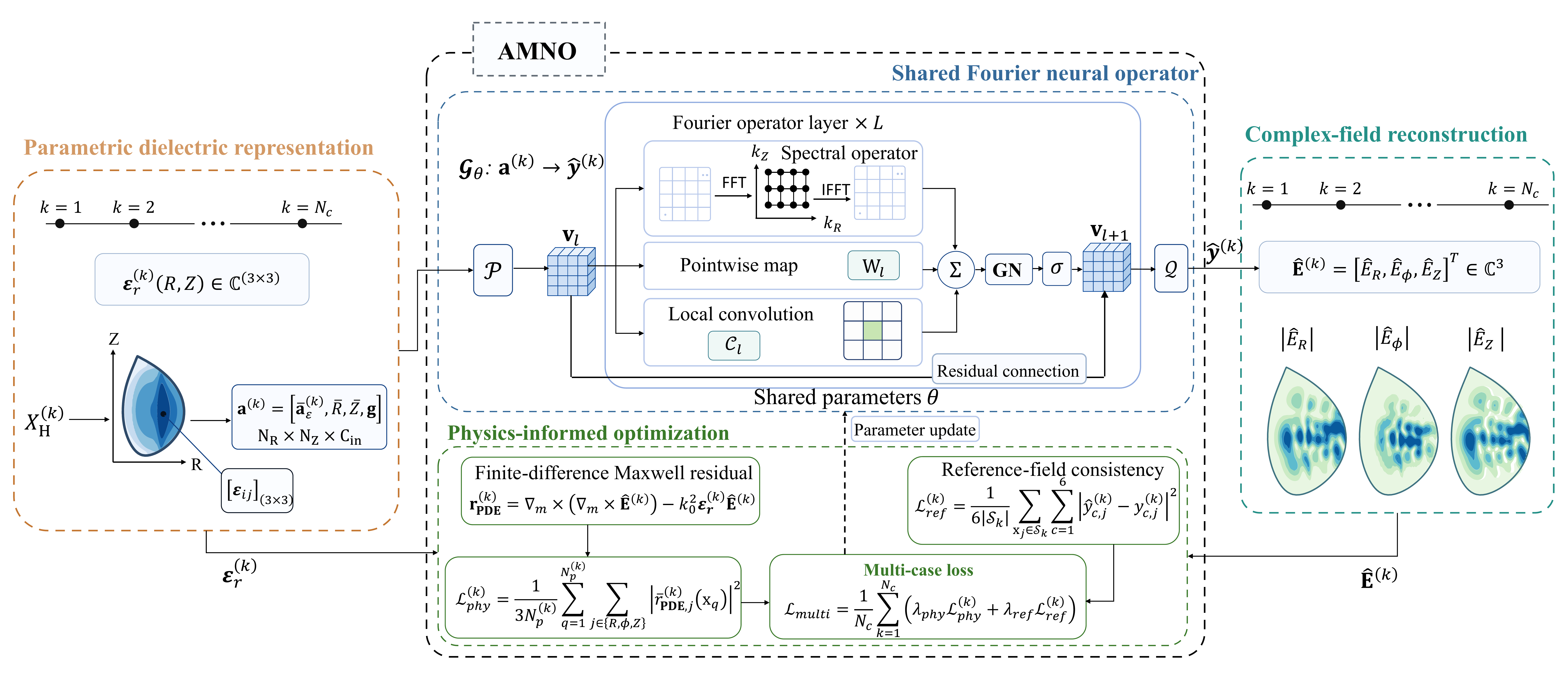}
\caption*{Figure 1. Overall architecture of AMNO. The spatially varying complex anisotropic dielectric-tensor field, spatial coordinates, and fixed problem encoding are mapped by a shared FNO to the three-component complex electric field. Training combines the frequency-domain Maxwell-equation residual and sparse reference-field consistency within a multi-case physics-informed objective.}
\end{figure*}

\section{Introduction}

Ion cyclotron resonance heating (ICRH) transfers radio-frequency energy to plasma particles by injecting high-power electromagnetic waves in the ion cyclotron range of frequencies (ICRF) into the plasma, and is one of the principal means of external power injection for high-temperature plasmas {[}1--6{]}. ICRH has been extensively investigated on the Experimental Advanced Superconducting Tokamak (EAST), including hydrogen-minority heating in deuterium plasmas and operation with a new ICRF antenna at 37 MHz {[}2,8{]}. Previous studies have shown that variations in plasma dielectric conditions can alter antenna coupling and the spatial distribution of the wave field {[}7--13{]}. Accurate determination of the spatial distributions of the three complex electric-field components under different plasma dielectric conditions therefore provides an important basis for analysing ICRH wave propagation and coupling processes and for performing subsequent power deposition calculations.

At present, ICRH electromagnetic responses are predominantly computed using full-wave numerical models. TORIC solves reduced finite-Larmor-radius wave equations using a mixed finite-element/spectral representation, whereas AORSA employs an all-orders spectral formulation in \(k\bot\rho_{i}\); both have been applied to ICRF mode-conversion problems {[}14--16{]}. These capabilities arise from the plasma-response formulations implemented in these dedicated ICRF solvers. By contrast, the COMSOL reference model adopted in the present work employs a pointwise local anisotropic dielectric closure and does not include the corresponding spatially dispersive FLR treatment or the associated linear mode-conversion physics. Integrated modelling further links full-wave fields, quasilinear dynamics, and power absorption {[}17,18{]}. For antenna-coupling problems and complex geometries, finite-element models such as RAPLICASOL have been validated and benchmarked against TOPICA, experimental results, and other full-wave solvers {[}19--21{]}. However, spatially varying complex dielectric tensors, non-Maxwellian distribution functions, the scrape-off layer (SOL), and full-wave--Fokker--Planck coupling substantially increase both model complexity and computational cost {[}22--25{]}. For parametric sweeps or predictions across multiple cases, conventional approaches generally require the repeated assembly and solution of large-scale discretised systems for each dielectric condition, making it difficult to reconcile computational accuracy with rapid-response requirements.

In recent years, physics-informed machine learning and neural operators have provided new approaches for the rapid solution of parametric partial differential equations (PDEs). Physics-informed neural networks (PINN) incorporate governing equations, initial and boundary conditions, and observational data into the loss function, thereby constraining network outputs by the underlying physical laws {[}26,27{]}. DeepONet and its physics-informed extensions further learn mappings from input functions to solution functions, enabling a single model to address a family of parametric equations {[}28,29{]}. The Fourier neural operator (FNO) uses nonlocal operator layers in Fourier space to represent global spatial coupling, while neural-operator theory emphasises the relative independence of the learned function-space mapping from discretisation resolution {[}30,31{]}. Using the same residual-based constraint principle as PINN, the physics-informed neural operator (PINO) combines data supervision with PDE residuals to learn a solution operator shared across a family of parametric PDEs, rather than a separate solution for an individual PDE instance {[}26,27,32,33{]}. Previous studies have also demonstrated that physical constraints, sparse observations, and physics-enhanced surrogate strategies can improve the reliability and generalisation performance of neural operators {[}34,35{]}.

In computational electromagnetics, MaxwellNet uses the residual of Maxwell's equations to constrain the mapping from a material-property distribution to the corresponding electric-field response {[}36{]}. Neural operators have also been applied to frequency-domain geophysical electromagnetic forward modelling, surrogate modelling of free-form electromagnetic scattering, and the solution of PDEs on general geometries {[}37--39{]}. In fusion research, a convolutional PINO has been used to reconstruct Grad--Shafranov equilibria and plasma separatrices {[}40{]}. In ICRF heating, Sánchez-Villar et al. developed real-time-capable surrogate models based on a TORIC simulation database for the expected operating scenarios of NSTX and WEST, using random forest and multilayer perceptron approaches. The models can rapidly predict multi-species core ICRF power absorption for high harmonic fast wave and ion cyclotron minority heating scenarios, while reducing the computational time by approximately six orders of magnitude, demonstrating their potential for integrated modelling and real-time control applications {[}41{]}. Existing studies have primarily addressed optical media, geophysical electromagnetic responses, or reduced-dimensional power quantities. To the best of our knowledge, the combination of sparse reference-field supervision and Maxwell-constrained operator learning has not previously been explored for mapping the complex anisotropic dielectric tensor of a magnetised tokamak plasma to its three-component complex electric-field response. Such an approach could provide a new route to solving for the three-component complex electric field in a unified manner under different plasma dielectric conditions and reduce repeated full-wave calculations required for parametric sweeps.

\begin{figure*}[!b]
\centering
\includegraphics[
  width=0.62\textwidth,
  height=0.58\textheight,
  keepaspectratio
]{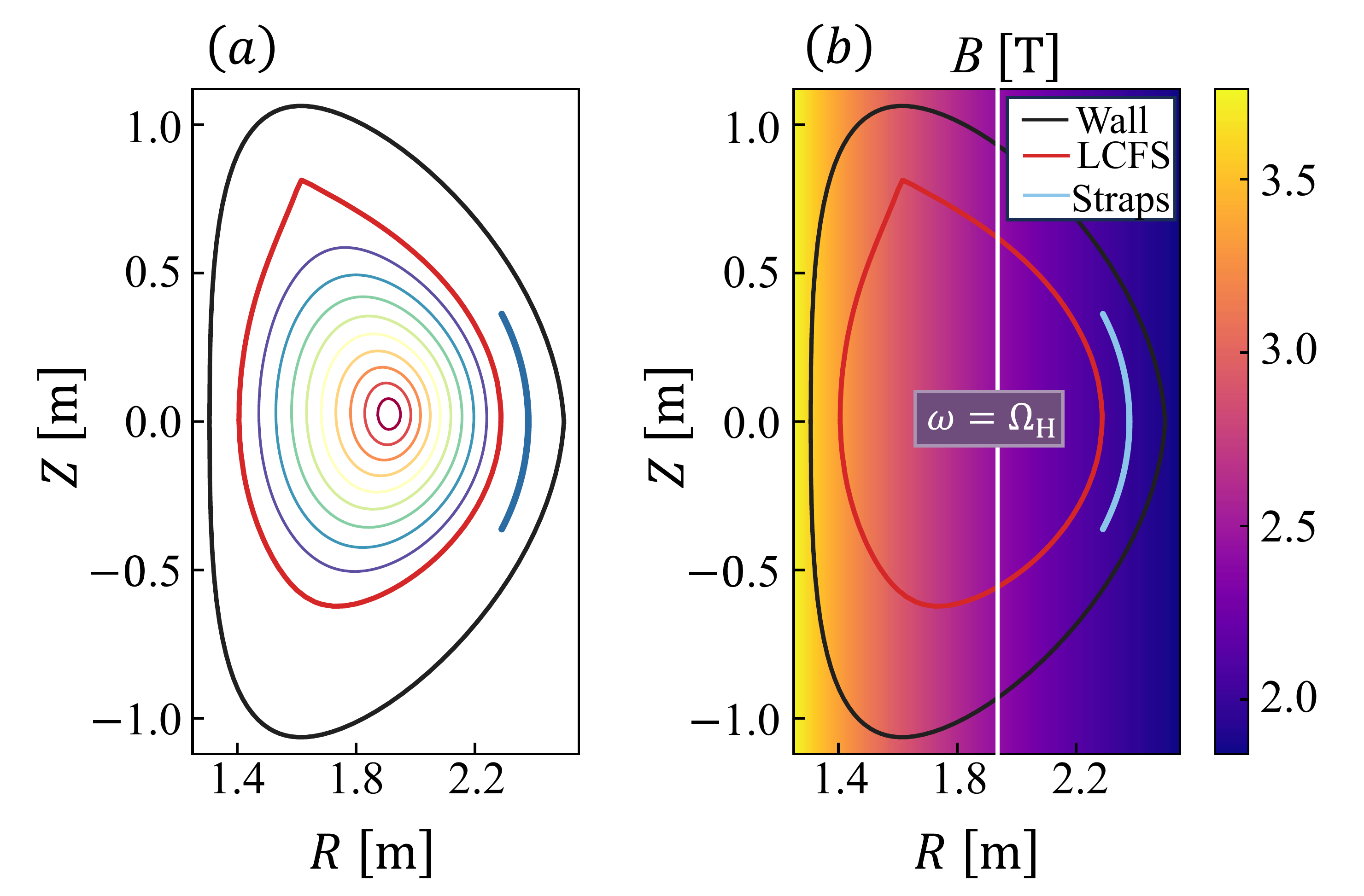}
\caption*{Figure 2. EAST poloidal configuration and magnetic-field profile adopted in the reference full-wave model. (a) Computational geometry showing the computational wall, selected closed magnetic surfaces, the last closed flux surface (LCFS), and the antenna straps. (b) Magnetic-field magnitude \(B = \frac{B_{0}R_{0}}{R}\), together with the fundamental hydrogen cyclotron resonance layer at 37 MHz.}
\end{figure*}

In this paper we propose an anisotropic Maxwell neural operator (AMNO) for rapid parametric full-wave modelling of ICRH. The proposed method reformulates the repeated full-wave numerical solutions required for the \(X_{H}\)-parameterised dielectric conditions considered in this study as a shared operator-learning problem that maps the corresponding complex anisotropic dielectric-tensor field to the three-component complex electric field, thereby establishing a physics-constrained surrogate framework for rapid in-range prediction within the modelled parameter family. With the operating frequency, toroidal mode number, and external excitation conditions held fixed, the hydrogen minority fraction is varied as the scanning parameter to construct a set of plasma cases with different dielectric conditions, and the corresponding spatially varying complex anisotropic dielectric-tensor field is calculated for each case. Spatial coordinates are additionally incorporated into the input, and AMNO jointly predicts the radial, toroidal, and vertical complex electric-field components. Through single-case complex-field reconstruction, shared multi-case training, and testing at two held-out \(X_{H}\) values within the modelled interval, the model is evaluated for its ability to reconstruct complex wave-field structures and spatial-scale composition across the \(X_{H}\)-generated dielectric conditions considered here. Comparisons with FNO baselines under sparse reference-field supervision are further conducted to isolate the contribution of the frequency-domain Maxwell-equation residual, while solution-time comparisons are used to assess the efficiency of parametric inference.


\section{Methods}

ICRH frequency-domain full-wave calculations under different plasma dielectric conditions can, in essence, be formulated as a family of Maxwell boundary-value problems parameterised by spatially varying complex anisotropic dielectric-tensor fields. Here, full-wave denotes the direct solution of the vector frequency-domain Maxwell boundary-value problem without a ray- or WKB-based approximation; the term does not imply a fully kinetic treatment of the plasma dielectric response. To enable a unified solution across multiple dielectric conditions, we construct AMNO with shared parameters to learn the operator mapping from the complex anisotropic dielectric-tensor field to the three-component complex electric field. The model takes the dielectric-tensor field, spatial coordinates, and fixed problem encoding as inputs, jointly predicts the radial, toroidal, and vertical complex electric-field components, and is constrained by both the frequency-domain Maxwell-equation residual and limited reference-field information. The overall framework of the proposed method is shown in figure 1.

\subsection{Frequency-domain ICRH model and input representation}

In the cylindrical coordinate system\((R,\phi,Z)\), we solve for the three-component complex electric field \(\widetilde{\mathbf{E}}(R,Z) = \left\lbrack {\widetilde{E}}_{R},{\widetilde{E}}_{\phi},{\widetilde{E}}_{Z} \right\rbrack^{T}\) in the tokamak \(R\text{--}Z\) poloidal cross-section at a prescribed RF angular frequency \(\omega\) and toroidal mode number \(m\). The following convention is adopted for the temporal and toroidal dependence:

\begin{equation*}
\begin{array}{r}
{\mathbf{E}(R,\phi,Z,t) = Re\ }\left\lbrack \widetilde{\mathbf{E}}(R,Z)\mathbb{e}^{i\omega t + im\phi} \right\rbrack,
\end{array}
\tag{1}
\end{equation*}

\(m\) is the signed toroidal mode number, so that \(\frac{\partial}{\partial_{t}} \rightarrow i\omega\), \(\frac{\partial}{\partial_{\phi}} \rightarrow im\). This expansion reduces the three-dimensional frequency-domain electromagnetic problem to a two-dimensional hybrid-mode problem on the \(R\text{--}Z\) cross-section. Nevertheless, \({\widetilde{E}}_{R}\),\({\widetilde{E}}_{\phi}\), and \({\widetilde{E}}_{Z}\) remain mutually coupled and must be solved simultaneously.

The permeability is set to the vacuum permeability. Equation (2) is applied in the source-free interior, where no impressed volume-current density is present, while the RF field is excited through a prescribed surface-current density on the antenna-strap boundary. Under these conditions, the frequency-domain Maxwell equation can be written as the following curl--curl wave equation for the electric field:

\begin{equation*}
\begin{array}{r}
\nabla_{m} \times \left( \nabla_{m} \times \widetilde{\mathbf{E}} \right) - k_{0}^{2}\boldsymbol{\varepsilon}_{r}\widetilde{\mathbf{E}} = \mathbf{0},\ \ k_{0} = \frac{\omega}{c_{0}},
\end{array}
\tag{2}
\end{equation*}

The quantity \(c_{0}\) is the speed of light in vacuum, \(\nabla_{m}\) denotes the modal differential operator obtained by replacing the toroidal derivative in the cylindrical-coordinate differential operators with \(\mathbb{i}m\), and \(\boldsymbol{\varepsilon}_{r}(R,Z) \in \mathbb{C}^{3 \times 3}\) is the spatially varying complex anisotropic relative dielectric tensor. Its off-diagonal elements describe the coupling among the electric-field components. In the present reference model, \(\boldsymbol{\varepsilon}_{r}(R,Z)\) is evaluated pointwise under a locally homogeneous plasma approximation. Thus, the spatially varying anisotropic dielectric response is retained, whereas nonlocal finite-Larmor-radius effects and the associated linear mode conversion are excluded. The outer-wall and antenna boundary conditions are identical to those adopted in the reference full-wave model.

The parametric cases are constructed by varying the hydrogen minority fraction, defined as

\begin{equation*}
\begin{array}{r}
X_{H} = \frac{n_{H}}{n_{e}},
\end{array}
\tag{3}
\end{equation*}

where \(n_{H}\) and \(n_{e}\) denote the number densities of the hydrogen minority ions and electrons, respectively. Under the two-ion-species H--D approximation adopted in this work, \(n_{D} = n_{e} - n_{H}\). For the \(k\)th case, the change in ion composition is first converted into the corresponding dielectric-tensor field, which subsequently determines the full-wave electric-field response:

\begin{equation*}
\begin{array}{r}
X_{H}^{(k)} \rightarrow \left\{ n_{H}^{(k)},n_{D}^{(k)} \right\} \rightarrow \boldsymbol{\varepsilon}_{\mathbf{r}}^{(k)}(R,Z) \rightarrow {\widetilde{\mathbf{E}}}^{(k)}(R,Z).
\end{array}
\tag{4}
\end{equation*}

The computational geometry, equilibrium magnetic field, radio frequency, toroidal mode number, antenna excitation, and other problem settings corresponding to the EAST configuration are held fixed across all cases; only the ion composition and its corresponding dielectric-tensor field are varied. The EAST poloidal configuration and magnetic-field profile adopted in the reference full-wave model are shown in figure 2.

The parameter \(X_{H}\) is used solely to generate the dielectric tensor, whereas AMNO directly learns the mapping \(\boldsymbol{\varepsilon}_{\mathbf{r}}^{(k)}(R,Z) \rightarrow {\widetilde{\mathbf{E}}}^{(k)}(R,Z)\).

The complex dielectric tensor for each case is mapped onto a common regular \(R\text{--}Z\) grid. The nine complex tensor components of the \(k\)th case are then vectorised in a prescribed order, and their real and imaginary parts are separated to obtain the dielectric feature field:

\begin{equation*}
\begin{array}{r}
\mathbf{a}_{\varepsilon}^{(k)} = \left\lbrack Re\ vec\left( \boldsymbol{\varepsilon}_{r}^{(k)}(R,Z) \right),Im\ vec\left( \boldsymbol{\varepsilon}_{r}^{(k)}(R,Z) \right) \right\rbrack,
\end{array}
\tag{5}
\end{equation*}

The operator \(vec( \cdot )\) denotes the expansion of the nine complex-valued components of the dielectric tensor into a channel vector according to a prescribed component order. Given the pronounced differences in numerical scale among the dielectric components, the dielectric feature channels are standardised consistently across cases:

\begin{equation*}
\begin{array}{r}
{\overline{a}}_{\varepsilon,c}^{(k)} = \frac{a_{\varepsilon,c}^{(k)} - \mu_{\varepsilon,c}}{\sigma_{\varepsilon,c}},
\end{array}
\tag{6}
\end{equation*}

Here, \(\mu_{\varepsilon,c}\) is the mean and \(\sigma_{\varepsilon,c}\) is the standard deviation of the corresponding channel over the valid grid points of the modelling cases. Finally, the standardised dielectric feature field is concatenated with the normalised spatial coordinates and the fixed problem encoding to form the input field of AMNO:

\begin{equation*}
\begin{array}{r}
\mathbf{a}^{(k)}(R,Z) = \left\lbrack {\overline{\mathbf{a}}}_{\varepsilon}^{(k)}(R,Z),\overline{R},\overline{Z},\mathbf{g}(R,Z) \right\rbrack\mathbf{,}
\end{array}
\tag{7}
\end{equation*}

In equation (7), \(\overline{R}\) and \(\overline{Z}\) are the normalised spatial coordinates, and \(\mathbf{g}(R,Z)\) denotes the fixed problem encoding shared across all cases.

\begin{figure*}[!b]
\centering
\includegraphics[width=0.73\textwidth,height=0.68\textheight,keepaspectratio]{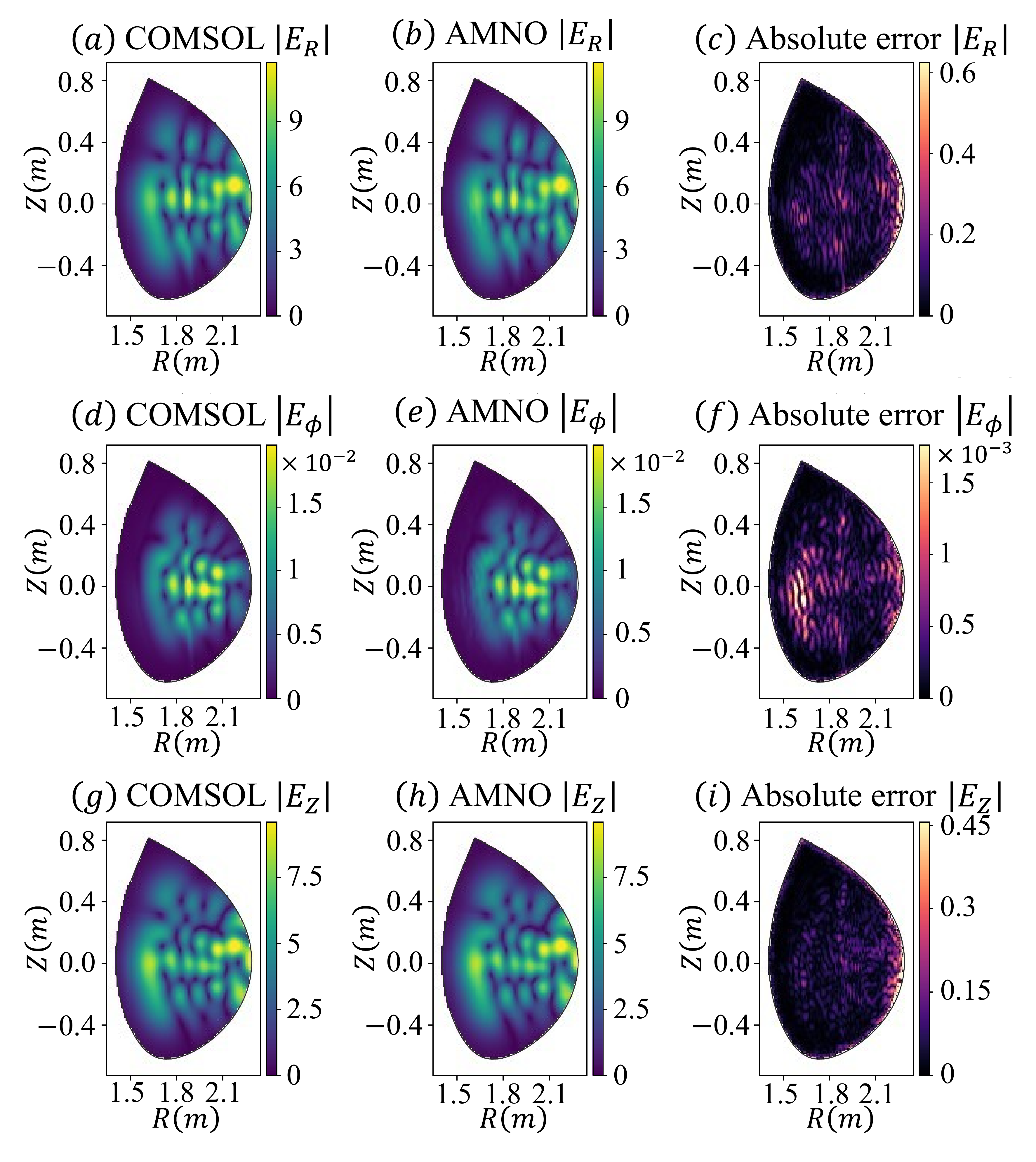}
\caption*{Figure 3. Predicted magnitudes of the three electric-field components in a representative case. The rows correspond to \(\left| E_{R} \right|\), \(\left| E_{\phi} \right|\), and \(\left| E_{Z} \right|\), respectively, while the columns show the COMSOL reference magnitudes, the AMNO predictions, and the absolute complex-field errors. For each electric-field component, the COMSOL and AMNO results use the same colour scale, whereas the error map uses an independent colour scale. All field quantities are expressed in \(V/m\).}
\end{figure*}

\begin{figure*}[!b]
\centering
\includegraphics[width=0.94\textwidth,height=0.68\textheight,keepaspectratio]{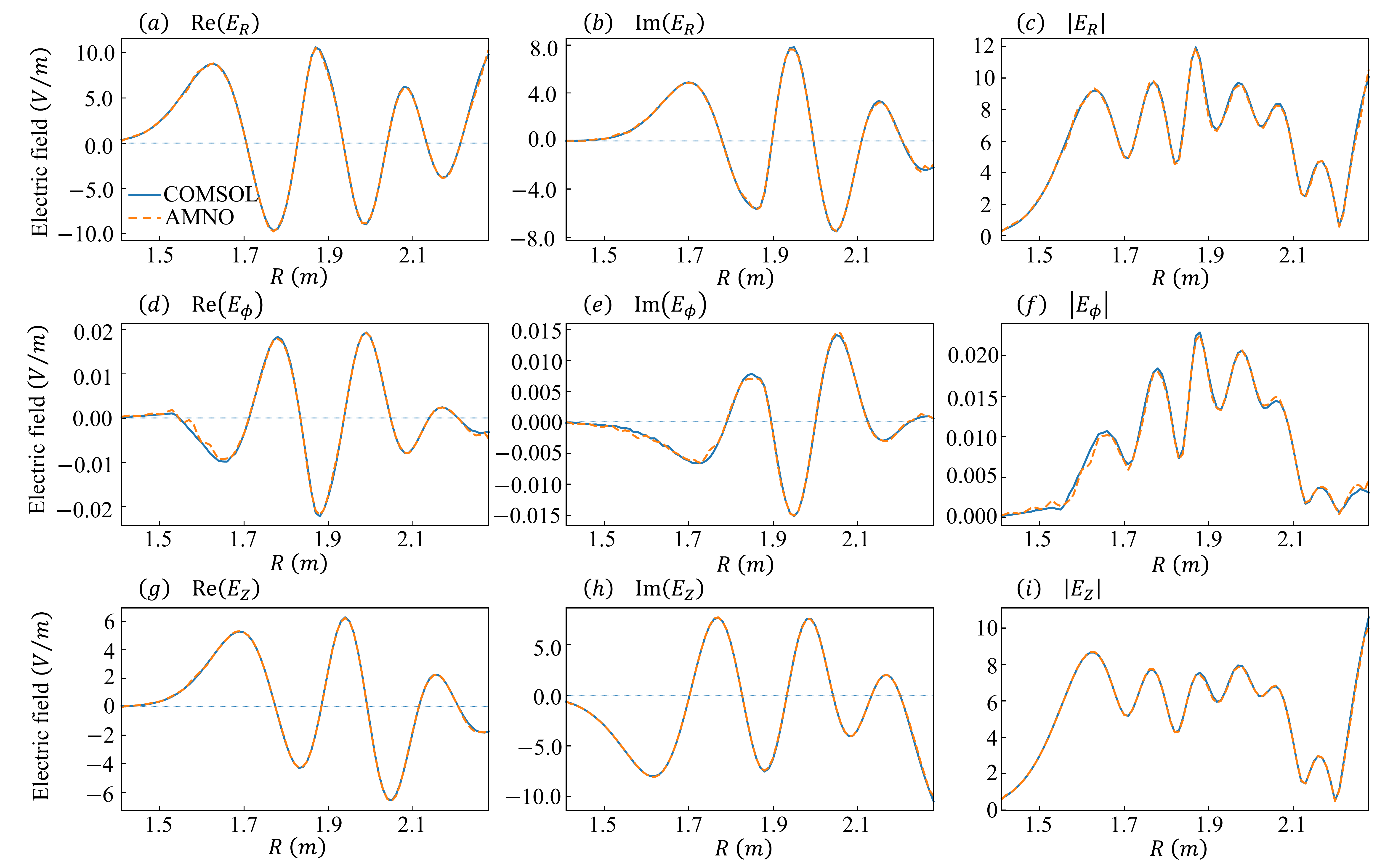}
\caption*{Figure 4. Comparison of the complex-electric-field line profiles along the \(R\) direction at the midplane (\(Z = 0\)) between the COMSOL reference solution and the AMNO prediction for a representative case. The rows correspond to \(E_{R}\), \(E_{\phi}\), and \(E_{Z}\), respectively, while the columns show the real part, imaginary part, and magnitude. All electric-field quantities are expressed in \(V/m\).}

\par\vspace{7pt}
\centering
\caption*{Table 1. Prediction errors of AMNO for the three complex electric-field components in the representative case.}
\begin{tabular}{@{}
  >{\centering\arraybackslash}m{(\linewidth - 6\tabcolsep) * \real{0.2269}}
  >{\centering\arraybackslash}m{(\linewidth - 6\tabcolsep) * \real{0.2554}}
  >{\centering\arraybackslash}m{(\linewidth - 6\tabcolsep) * \real{0.2556}}
  >{\centering\arraybackslash}m{(\linewidth - 6\tabcolsep) * \real{0.2556}}@{}}
\toprule\noalign{}
Electric-field component & \(MSE(V/m)^{2}\) & \(MAE(V/m)\) & Relative \(L_{2}\) error \\
\midrule\noalign{}
\(E_{R}\) & \(1.56 \times 10^{- 2}\) & \(7.82 \times 10^{- 2}\) & \(2.87 \times 10^{- 2}\) \\
\(E_{\phi}\) & \(1.75 \times 10^{- 7}\) & \(2.77 \times 10^{- 4}\) & \(6.96 \times 10^{- 2}\) \\
\(E_{Z}\) & \(7.62 \times 10^{- 3}\) & \(5.04 \times 10^{- 2}\) & \(2.35 \times 10^{- 2}\) \\
\bottomrule\noalign{}
\end{tabular}
\end{figure*}

\begin{figure*}[!b]
\centering
\includegraphics[width=0.75\textwidth,height=0.68\textheight,keepaspectratio]{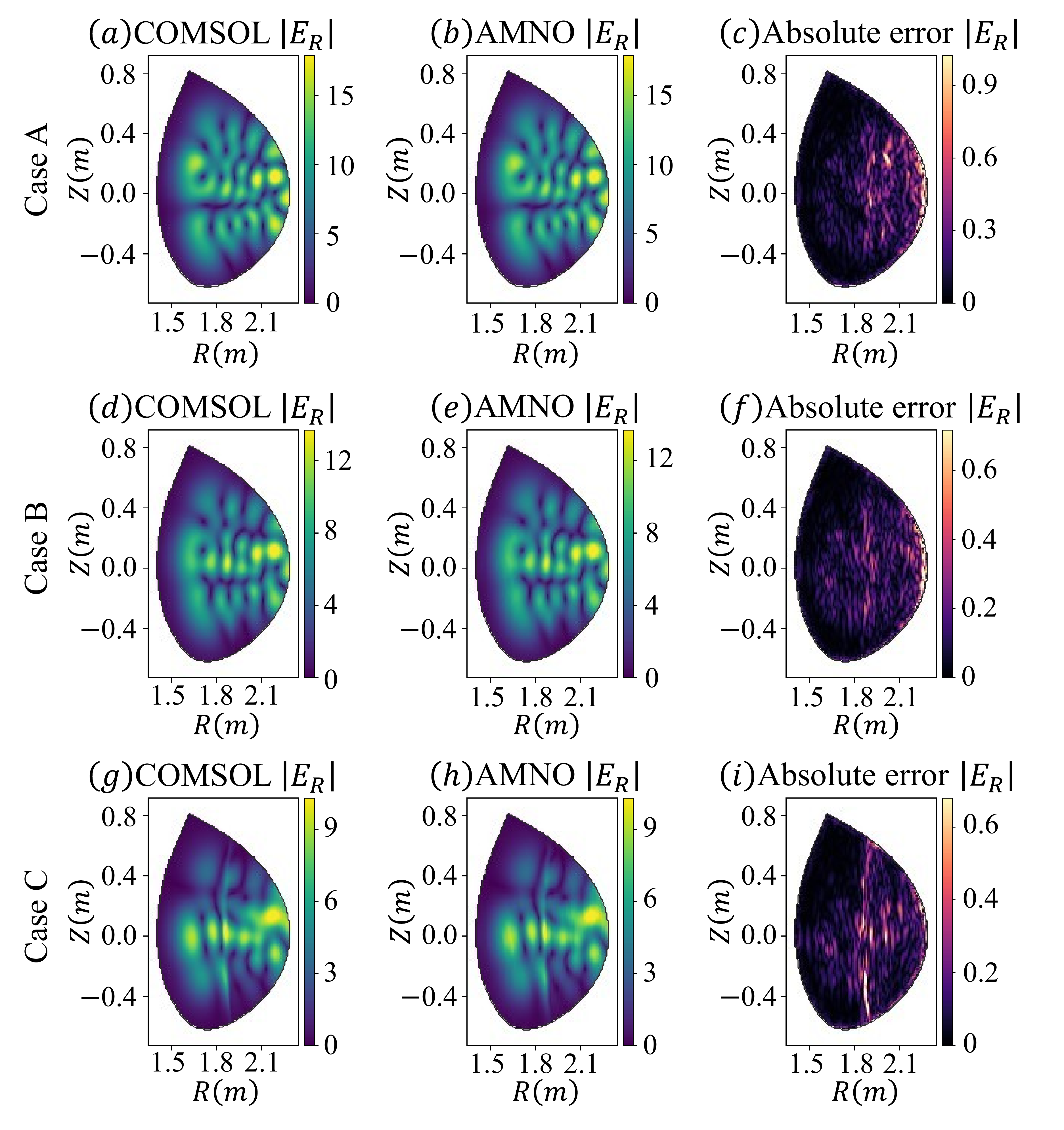}
\caption*{Figure 5. Predicted magnitudes of the radial electric-field component \(E_{R}\) for modelling cases A--C. The rows correspond to cases A, B, and C, respectively, while the columns show the COMSOL reference results, the AMNO predictions, and the corresponding absolute complex-field errors. All field quantities are expressed in \(V/m\).}
\end{figure*}

\begin{figure*}[b]
\centering
\includegraphics[width=1.00\textwidth,height=0.68\textheight,keepaspectratio]{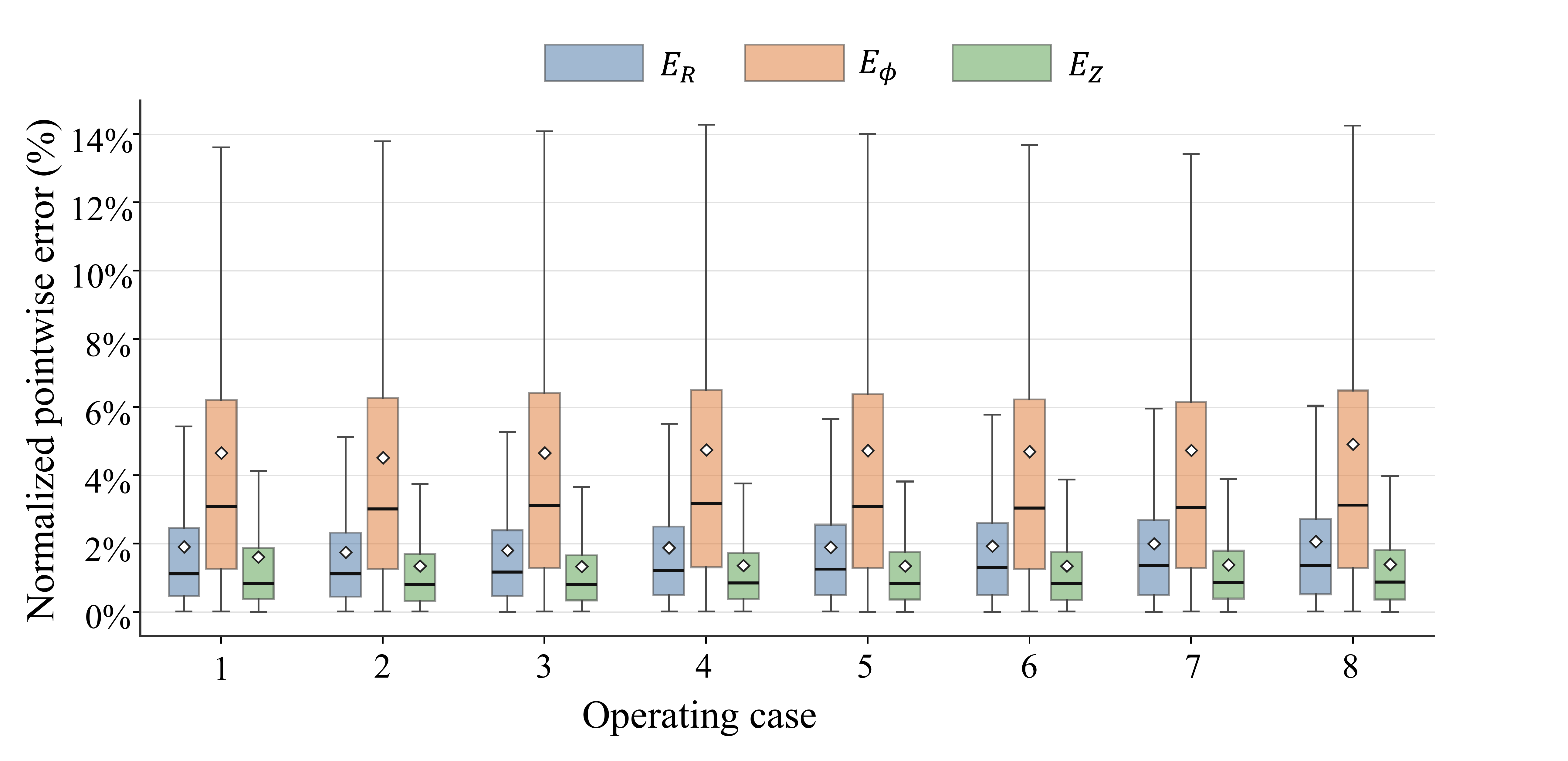}
\caption*{Figure 6. Distributions of the RMS-normalised pointwise errors of the three complex electric-field components for eight modelling cases. Cases 1--8 are ordered by increasing hydrogen minority fraction \(X_{H}\) over the range \(0.01\sim 0.05\). The blue, orange, and green boxes correspond to \(E_{R}\), \(E_{\phi}\), and \(E_{Z}\), respectively. Each box spans the interquartile range from the 25th to the 75th percentile; the horizontal line within the box denotes the median, the open diamond denotes the mean, and the whiskers indicate the range of the error distribution.}
\end{figure*}

\begin{figure*}[b]
\centering
\includegraphics[width=1.00\textwidth,height=0.68\textheight,keepaspectratio]{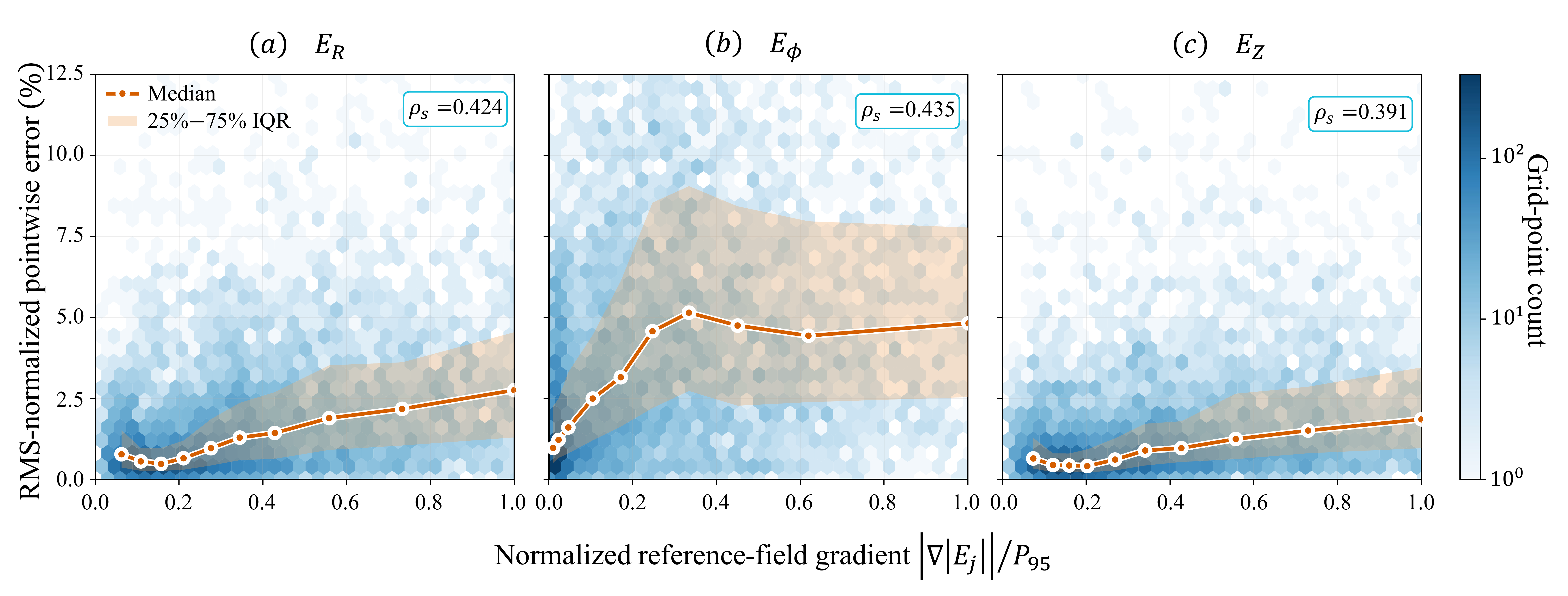}
\caption*{Figure 7. Statistical relationship between the local gradient of the reference-field magnitude and the pointwise complex-field error of AMNO for case A. Panels (a)--(c) correspond to \(E_{R}\), \(E_{\phi}\), and \(E_{Z}\), respectively. The horizontal axis represents the gradient of the reference-field magnitude normalised by its 95th percentile, and the vertical axis represents the pointwise complex-field error normalised by the root-mean-square value of the corresponding reference field. The colour of each hexagon indicates the number of grid points in the corresponding bin within the last closed flux surface and is displayed on a logarithmic scale. The solid line with circular markers and the shaded band represent, respectively, the median and the 25th--75th percentile range of the error within 10 equal-count gradient bins. \(\rho_{s}\) denotes the Spearman rank correlation coefficient calculated using all valid grid points.}
\end{figure*}

\begin{figure*}[t]
\centering
\includegraphics[width=0.82\textwidth,height=0.68\textheight,keepaspectratio]{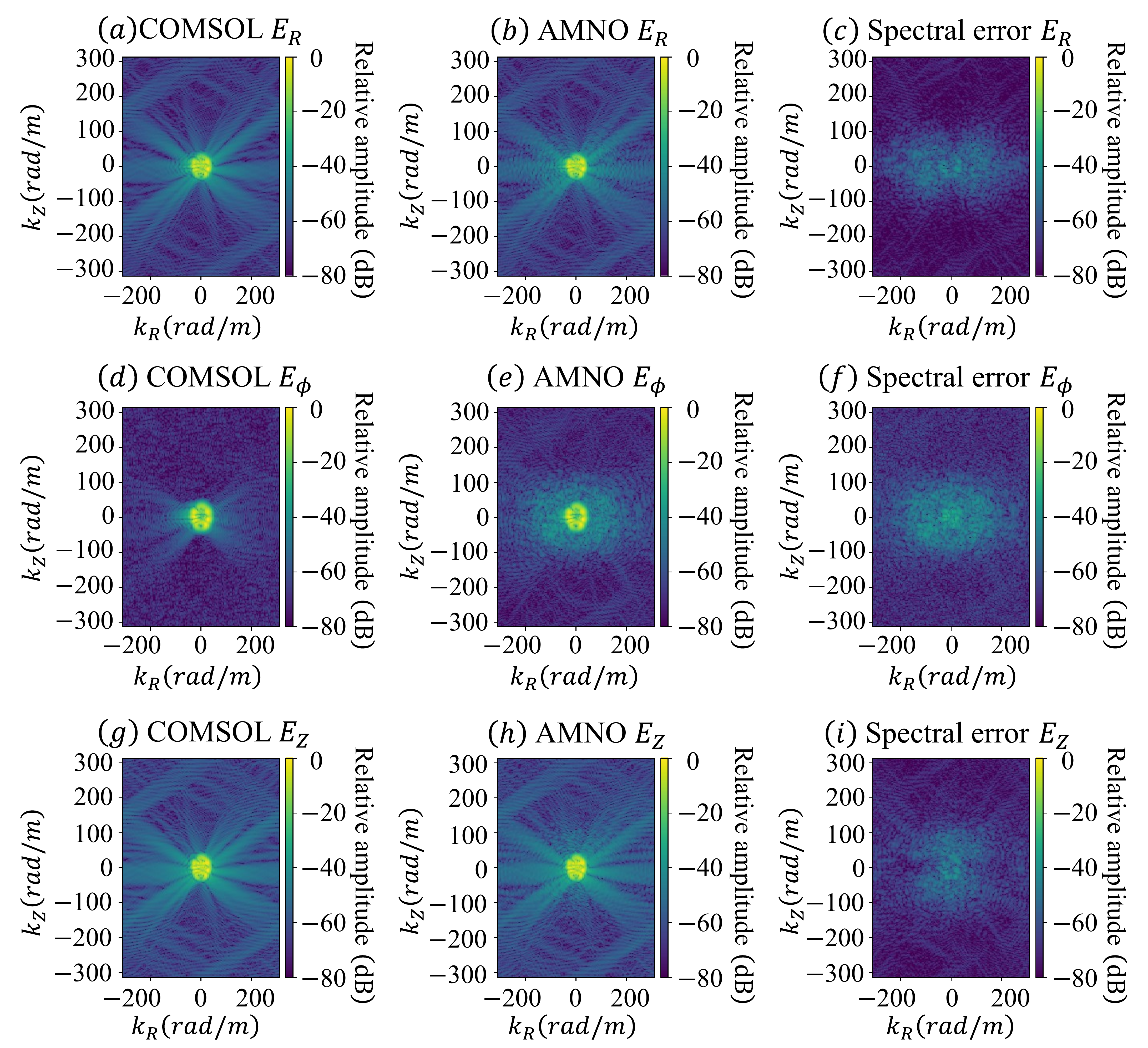}
\caption*{Figure 8. Two-dimensional spatial spectra and spectral errors of the three complex electric-field components for case A. The rows correspond to \(E_{R}\), \(E_{\phi}\), and \(E_{Z}\), respectively, while the columns show the COMSOL reference spectra, the AMNO predicted spectra, and the corresponding spectral errors. Both the spectral magnitudes and the spectral errors are normalised by the maximum spectral magnitude of the corresponding COMSOL component and expressed in dB.}
\end{figure*}

\begin{figure*}[t]
\centering
\includegraphics[width=0.70\textwidth,height=0.68\textheight,keepaspectratio]{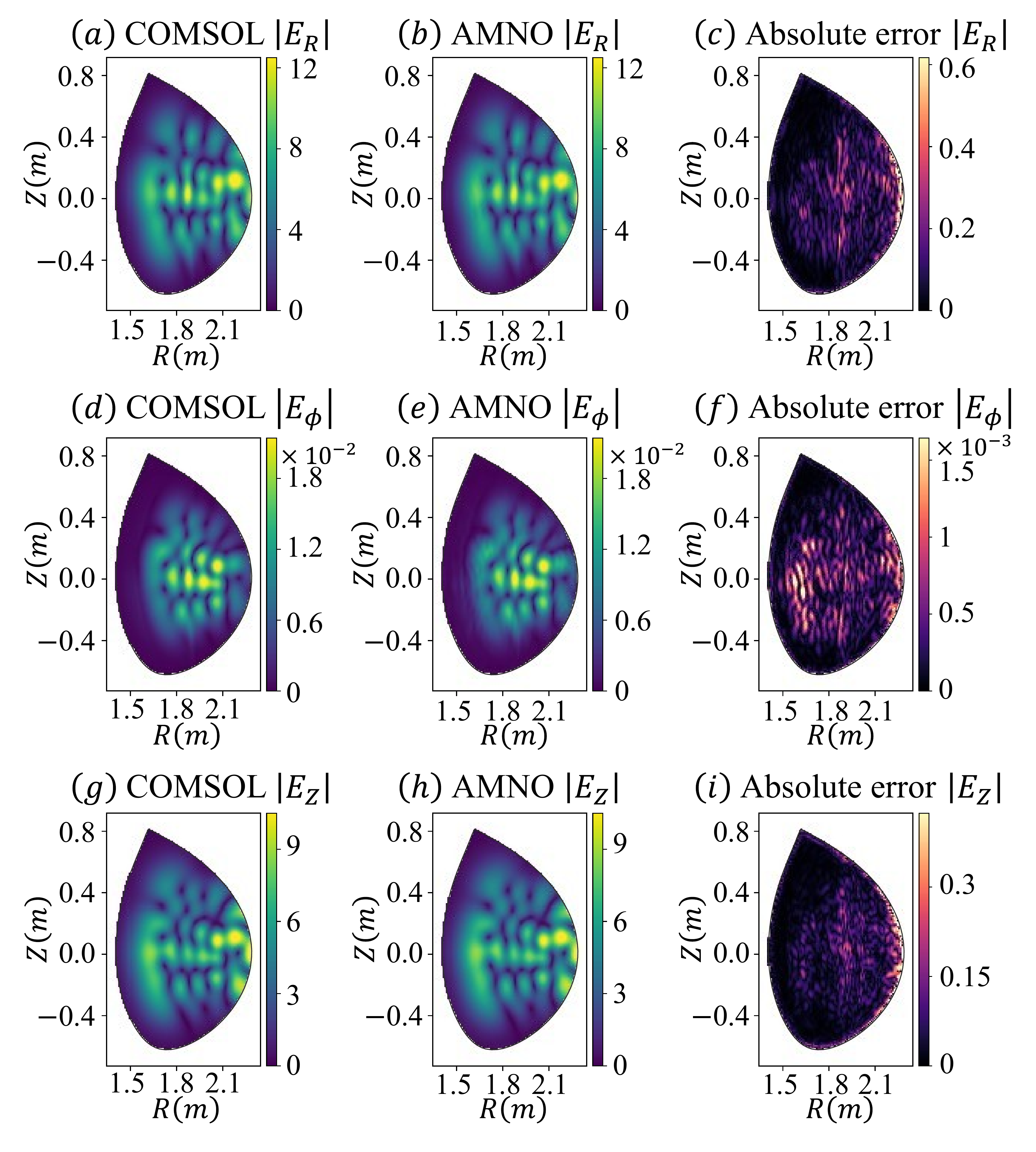}
\caption*{Figure 9. Predicted magnitudes of the three electric-field components for test case D. The rows correspond to \(\left| E_{R} \right|\), \(\left| E_{\phi} \right|\) and \(\left| E_{Z} \right|\), respectively, while the columns show the COMSOL reference magnitudes, the AMNO predicted magnitudes, and the absolute complex-field errors. All field quantities are expressed in \(V/m\).}
\end{figure*}

\begin{figure*}[t]
\centering
\includegraphics[width=0.70\textwidth,height=0.68\textheight,keepaspectratio]{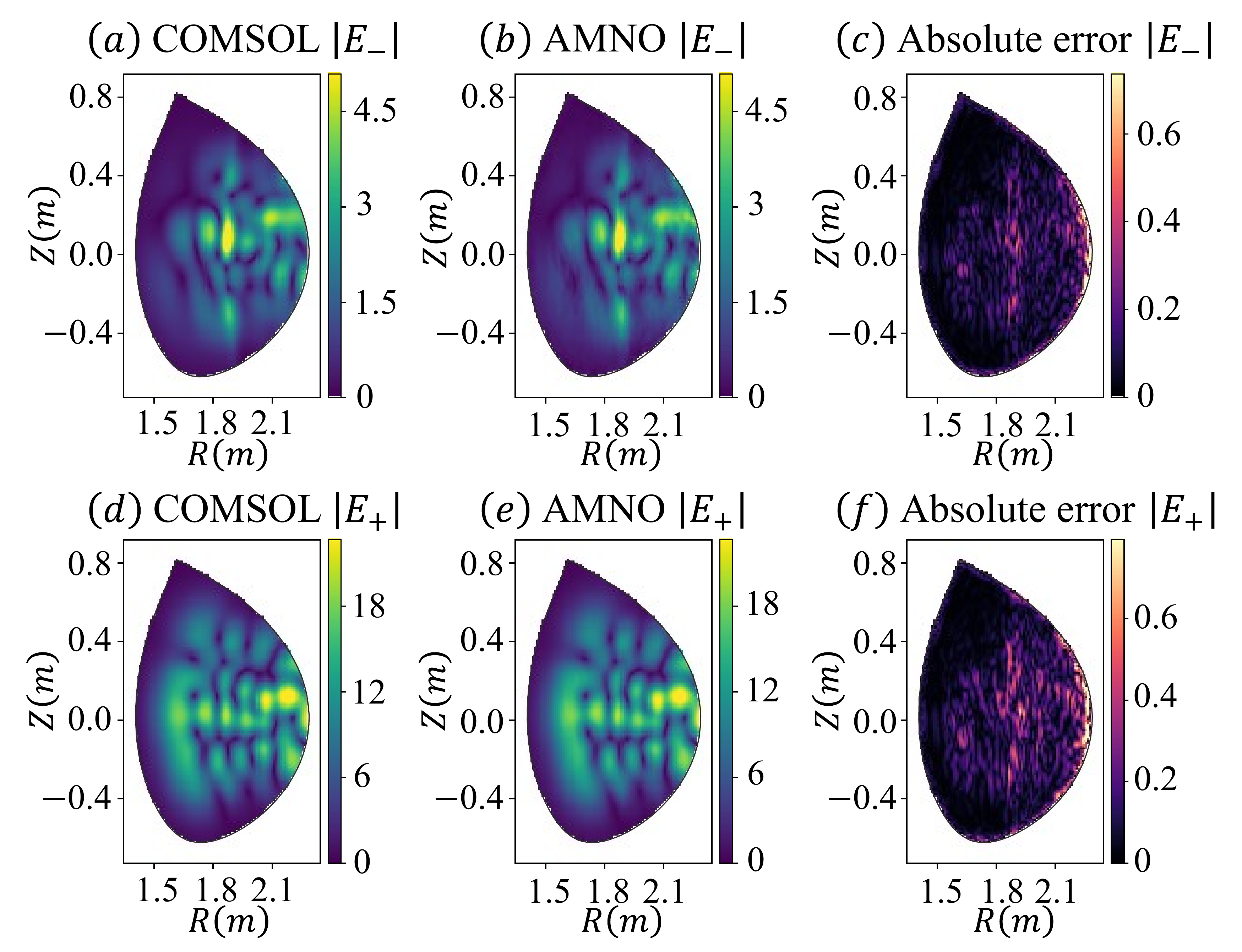}
\caption*{Figure 10. Reconstructed magnitudes of the polarisation-combination fields for test case D. The upper and lower rows correspond to \(E_{-}\) and \(E_{+}\), respectively, while the columns show the COMSOL reference results, the AMNO reconstructed results, and the absolute errors. The COMSOL and AMNO results use the same colour scale for each polarisation-combination field, whereas the error map uses an independent colour scale. All field quantities are expressed in \(V/m\).}

\par\vspace{7pt}
\centering
\caption*{Table 2. Prediction errors of AMNO for the three complex electric-field components in unseen interpolation test Cases D and E.}
\begin{tabular}{@{}
  >{\centering\arraybackslash}m{(\linewidth - 8\tabcolsep) * \real{0.1148}}
  >{\centering\arraybackslash}m{(\linewidth - 8\tabcolsep) * \real{0.1803}}
  >{\centering\arraybackslash}m{(\linewidth - 8\tabcolsep) * \real{0.2349}}
  >{\centering\arraybackslash}m{(\linewidth - 8\tabcolsep) * \real{0.2350}}
  >{\centering\arraybackslash}m{(\linewidth - 8\tabcolsep) * \real{0.2350}}@{}}
\toprule\noalign{}
Case & Electric-field component & \(MSE(V/m)^{2}\) & \(MAE(V/m)\) & Relative \(L_{2}\) error \\
\midrule\noalign{}
\multirow{3}{=}{\centering\arraybackslash Case D} & \(E_{R}\) & \(1.82 \times 10^{- 2}\) & \(8.82 \times 10^{- 2}\) & \(2.87 \times 10^{- 2}\) \\
& \(E_{\phi}\) & \(1.99 \times 10^{- 7}\) & \(3.09 \times 10^{- 4}\) & \(6.84 \times 10^{- 2}\) \\
& \(E_{Z}\) & \(7.78 \times 10^{- 3}\) & \(5.50 \times 10^{- 2}\) & \(2.18 \times 10^{- 2}\) \\
\midrule\noalign{}
\multirow{3}{=}{\centering\arraybackslash Case E} & \(E_{R}\) & \(1.40 \times 10^{- 2}\) & \(7.95 \times 10^{- 2}\) & \(2.93 \times 10^{- 2}\) \\
& \(E_{\phi}\) & \(1.49 \times 10^{- 7}\) & \(2.60 \times 10^{- 4}\) & \(7.02 \times 10^{- 2}\) \\
& \(E_{Z}\) & \(5.22 \times 10^{- 3}\) & \(4.64 \times 10^{- 2}\) & \(2.13 \times 10^{- 2}\) \\
\bottomrule\noalign{}
\end{tabular}

\par\vspace{7pt}
\centering
\caption*{Table 3. Comparison of the relative \(L_{2}\) errors of the three complex electric-field components between the modelling and test cases. The modelling-case results are reported as the mean \(\pm\) inter-case standard deviation, while Cases D and E give the corresponding individual test-case errors.}
\begin{tabular}{@{}
  >{\centering\arraybackslash}m{(\linewidth - 6\tabcolsep) * \real{0.1885}}
  >{\centering\arraybackslash}m{(\linewidth - 6\tabcolsep) * \real{0.2512}}
  >{\centering\arraybackslash}m{(\linewidth - 6\tabcolsep) * \real{0.2512}}
  >{\centering\arraybackslash}m{(\linewidth - 6\tabcolsep) * \real{0.2512}}@{}}
\toprule\noalign{}
Electric-field component & \({\overline{\varepsilon}}_{train} \pm \sigma_{train}\) & \(Case\ D\ \left( L_{2} \right)\) & \(Case\ E\ \left( L_{2} \right)\) \\
\midrule\noalign{}
\(E_{R}\) & \((2.90 \pm 0.10) \times 10^{- 2}\) & \(2.87 \times 10^{- 2}\) & \(2.93 \times 10^{- 2}\) \\
\(E_{\phi}\) & \((6.91 \pm 0.26) \times 10^{- 2}\) & \(6.84 \times 10^{- 2}\) & \(7.02 \times 10^{- 2}\) \\
\(E_{Z}\) & \((2.26 \pm 0.25) \times 10^{- 2}\) & \(2.18 \times 10^{- 2}\) & \(2.13 \times 10^{- 2}\) \\
\bottomrule\noalign{}
\end{tabular}
\end{figure*}

\begin{figure*}[t]
\centering
\includegraphics[width=0.67\textwidth,height=0.68\textheight,keepaspectratio]{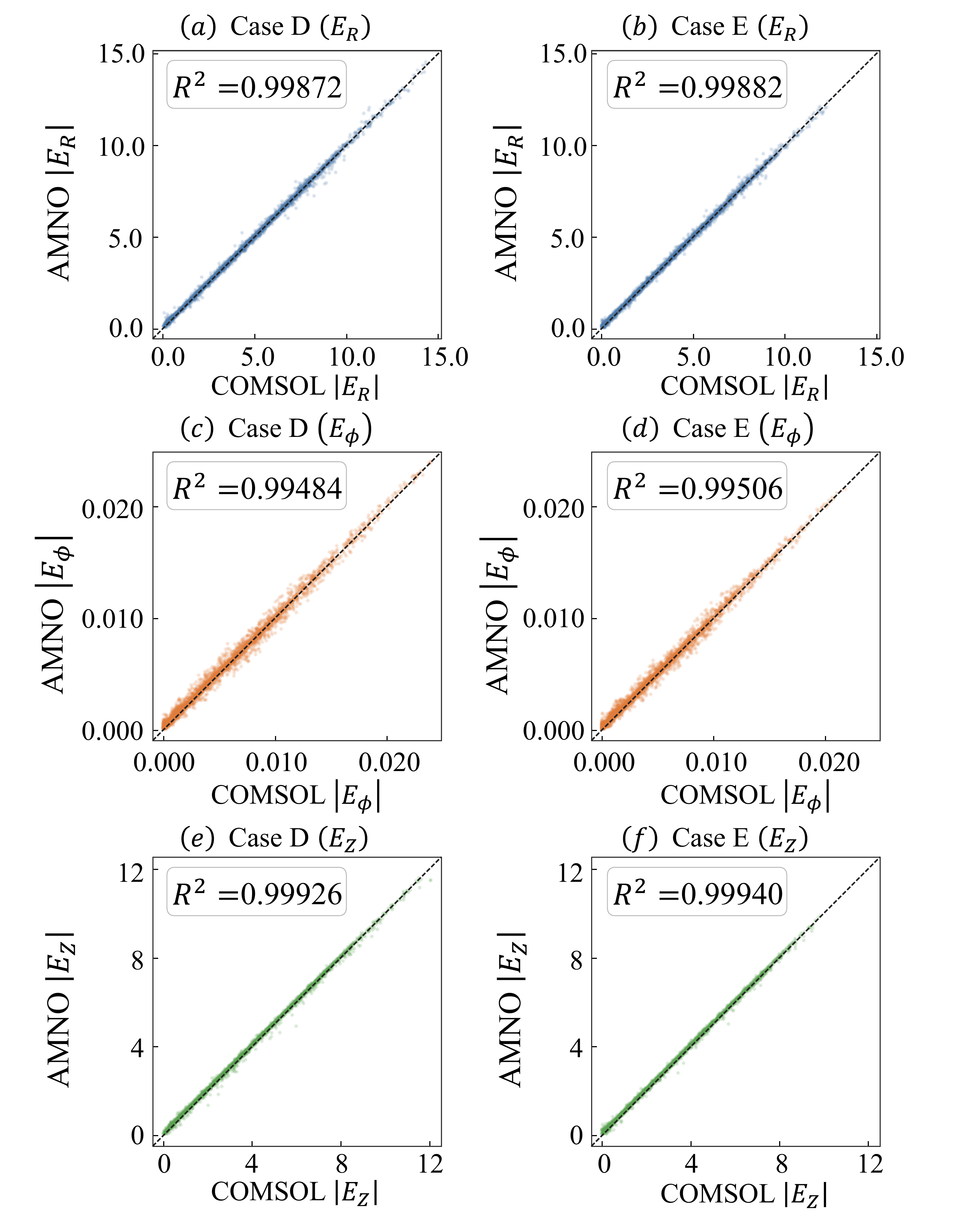}
\caption*{Figure 11. COMSOL--AMNO parity plots for the magnitudes of the three electric-field components in test cases D and E. The left and right columns correspond to Case D and Case E, respectively, while the upper, middle, and lower rows correspond to \(\left| E_{R} \right|\), \(\left| E_{\phi} \right|\) and \(\left| E_{Z} \right|\). The dashed line \(y = x\) represents the ideal prediction. The plotted points comprise 5000 points obtained by deterministic downsampling from all valid grid points, whereas the coefficient of determination \(R^{2}\) is calculated using all valid grid points. All coordinate axes are expressed in \(V/m\).}
\end{figure*}

\subsection{Multi-case complex-field operator learning and physics-constrained training of AMNO}

Using the multichannel input constructed in equation (7), AMNO learns a shared solution operator over the modelling cases belonging to the \(X_{H}\)-generated dielectric-field family, mapping each corresponding dielectric-tensor field to its three-component complex electric field. For the \emph{k}th case, the operator mapping is expressed as

\begin{equation*}
\begin{array}{r}
{\widehat{\mathbf{y}}}^{(k)} = \mathcal{G}_{\theta}\left( \mathbf{a}^{(k)} \right),
\end{array}
\tag{8}
\end{equation*}

where \(\mathbf{a}^{(k)}\) is the multichannel input field constructed in equation (7), and \({\widehat{\mathbf{y}}}^{(k)}\) is the real-valued representation of the three-component complex electric field output by the network, comprising the real and imaginary parts of \(E_{R}\), \(E_{\phi}\), and \(E_{Z}\). Here, \(\mathcal{G}_{\theta}\) denotes AMNO parameterised by \(\theta\).

The ICRH wave field contains both long-range spatial coupling across the poloidal cross-section and local oscillations and fine-scale lobe structures. Accordingly, AMNO employs a Fourier neural operator backbone comprising a global mapping in the spectral domain, pointwise channel transformations, and local spatial convolutions. The network first maps the original input channels into a high-dimensional latent feature space of width \(w\) through the lifting layer \(\mathcal{P}\). The resulting features are then successively processed by \(L\) Fourier operator layers to extract spatial coupling features, before being decoded into the predicted field by the projection layer \(\mathcal{Q}\):

\begin{equation*}
\begin{array}{r}
\mathbf{v}_{0}\mathcal{= P}\left( \mathbf{a}^{(k)} \right),\ \ {\widehat{\mathbf{y}}}^{(k)}\mathcal{= Q}\left( \mathbf{v}_{L} \right).
\end{array}
\tag{9}
\end{equation*}

\(\mathbf{v}_{l}\) denotes the latent feature field at layer \(l\). The \(l\)th Fourier operator layer is expressed as

\begin{equation*}
\begin{aligned}
\mathbf{v}_{l + 1} ={}& \mathbf{v}_{l} + \sigma\Bigl\{ \mathcal{N}_{l}\Bigl[
\mathcal{F}^{- 1}\left( \mathcal{R}_{l}\mathcal{ F}\left( v_{l} \right) \right)\\
&\qquad{}+ W_{l}\mathbf{v}_{l} + \mathcal{C}_{l}\left( \mathbf{v}_{l} \right) \Bigr] \Bigr\},\quad l = 0,...,L - 1,
\end{aligned}
\tag{10}
\end{equation*}

Here, \(\mathcal{F}\) and \(\mathcal{F}^{- 1}\) denote the two-dimensional Fourier transform and inverse Fourier transform, respectively;\allowbreak\(\mathcal{R}_{l}\) is a learnable complex-valued mapping on the retained Fourier modes;\allowbreak\(W_{l}\) is a pointwise linear map; \(\mathcal{C}_{l}\) is a local spatial convolution; \(\mathcal{N}_{l}\) denotes Group Normalization (GN); and \(\sigma\) is a nonlinear activation function.

The spectral branch establishes nonlocal spatial connections through learnable complex-valued mappings on the retained Fourier modes. The pointwise map mixes information across the latent feature channels, whereas the local convolutional branch supplements fine-scale information from neighbouring regions. After normalisation and nonlinear transformation, the three types of features form a residual connection with the layer input \(\mathbf{v}_{l}\), thereby accounting for both global spatial coupling and local structural representation.

Because the network operates on real-valued tensors, the three-component complex electric field is encoded into six real-valued channels and jointly decoded from the shared latent features:

\begin{equation*}
\begin{aligned}
\mathbf{y}^{(k)} ={}& \left\lbrack Re\ E_{R}^{(k)},Im\ E_{R}^{(k)},Re\ E_{\phi}^{(k)},\right.\\
&\left.{}Im\ E_{\phi}^{(k)},Re\ E_{Z}^{(k)},Im\ E_{Z}^{(k)} \right\rbrack,
\end{aligned}
\tag{11}
\end{equation*}

\(E_{R}\), \(E_{\phi}\), and \(E_{Z}\) differ substantially in amplitude scale; direct joint optimisation at their original physical scales would cause the large-amplitude components to dominate the loss function and the backpropagated gradients. To balance the contributions of the different components, a separate component scale \(s_{E,j}\) is introduced for each electric-field component, and the three complex electric-field components are normalised as

\begin{equation*}
\begin{array}{r}
{\overline{E}}_{j} = \frac{E_{j}}{s_{E,j}},\ \ j \in \left\{ R,\phi,Z \right\},
\end{array}
\tag{12}
\end{equation*}

The component scale \(s_{E,j}\) is determined from the statistical scale of the corresponding electric-field component over the modelling cases. The network outputs six real-valued channels in the normalised space. Their physical scales are subsequently restored, and the corresponding real and imaginary parts are recombined to reconstruct the complex electric field:

\begin{equation*}
\begin{array}{r}
{\widehat{E}}_{j} = s_{E,j}\left( {\widehat{\overline{E}}}_{j,r} + i{\widehat{\overline{E}}}_{j,i} \right),\ \ j \in \left\{ R,\phi,Z \right\}.
\end{array}
\tag{13}
\end{equation*}

Unlike a purely data-driven operator, AMNO uses the frequency-domain Maxwell equation to construct a physical constraint that acts directly on the network output. The six real-valued channels predicted by the network are first reconstructed into the three-component complex electric field using equation (13), and the reconstructed field is then substituted into equation (2) to calculate the governing-equation residual:

\begin{equation*}
\begin{array}{r}
\mathbf{r}_{\mathbf{PDE}}^{(k)} = \nabla_{m} \times \left( \nabla_{m} \times {\widehat{\mathbf{E}}}^{(k)} \right) - k_{0}^{2}\boldsymbol{\varepsilon}_{\mathbf{r}}^{(k)}{\widehat{\mathbf{E}}}^{(k)}.
\end{array}
\tag{14}
\end{equation*}

Here, \(\nabla_{m}\) is the same modal differential operator as that defined in section 2.1. Because the FNO outputs the complete field on a regular \(R\text{--}Z\) grid in a single forward pass, the spatial derivatives are evaluated using finite differences on the regular grid. The finite-difference operations are first applied to the complete predicted field, after which the residual is aggregated only over the valid grid points included in the physical constraint. To reduce the scale differences among the residual components, the real and imaginary parts of the three complex residual components are normalised separately. The physics loss for the \(k\)th case is then defined as

\begin{equation*}
\begin{aligned}
\mathcal{L}_{phy}^{(k)} ={}& \frac{1}{3N_{p}^{(k)}}\sum_{q = 1}^{N_{p}^{(k)}}
\sum_{j \in \left\{ R,\phi,Z \right\}}^{}\Bigl\lbrack
\left| Re\ {\overline{\mathbf{r}}}_{\mathbf{PDE},j}^{(k)}\left( x_{q} \right) \right|^{2}\\
&\qquad{}+ \left| Im\ {\overline{\mathbf{r}}}_{\mathbf{PDE},j}^{(k)}\left( x_{q} \right) \right|^{2} \Bigr\rbrack,
\end{aligned}
\tag{15}
\end{equation*}

\(N_{p}^{(k)}\) is the number of physics grid points used in the residual calculation, and \({\overline{\mathbf{r}}}_{\mathbf{PDE},j}^{(k)}\) is the normalised \(j\)th residual component. Because the homogeneous governing equation within the domain cannot uniquely determine the non-trivial response associated with the prescribed excitation, a reference-field consistency loss is further defined over a sparse set of reference points \(\mathcal{S}_{k}\):

\begin{equation*}
\begin{array}{r}
\mathcal{L}_{ref}^{(k)} = \frac{1}{6\left| \mathcal{S}_{k} \right|}\sum_{x_{j} \in \mathcal{S}_{k}}^{}{\sum_{c = 1}^{6}\left| {\widehat{y}}_{c,j}^{(k)} - y_{c,j}^{(k)} \right|^{2}},
\end{array}
\tag{16}
\end{equation*}

Here, \(c\) indexes the six normalised real-valued channels of the complex electric field. The combined objective for the \(k\)th case is

\begin{equation*}
\begin{array}{r}
\mathcal{L}^{(k)} = \lambda_{phy}\mathcal{L}_{phy}^{(k)} + \lambda_{ref}\mathcal{L}_{ref}^{(k)},
\end{array}
\tag{17}
\end{equation*}

\(\lambda_{phy}\) and \(\lambda_{ref}\) are the weights assigned to the physics loss and the reference-field loss, respectively. For a training set containing \(N_{c}\) modelling cases, the shared operator parameters are optimised according to

\begin{equation*}
\begin{array}{r}
\theta^{*} = \arg\min_{\theta}\mathcal{L}_{multi}(\theta),\ \ \mathcal{L}_{multi} = \frac{1}{N_{c}}\sum_{k = 1}^{N_{c}}\mathcal{L}^{(k)},
\end{array}
\tag{18}
\end{equation*}

Complex-field reconstruction, finite-difference residual calculation, and the combined loss are all embedded in the computational graph used for network training. The loss gradients are backpropagated from the output to the projection layer and the Fourier operator layers, thereby updating the parameters \(\theta\) shared by all cases. Through this offline operator-learning process, independent case-by-case solutions are recast as a unified optimisation problem with shared parameters.

\begin{figure*}[!t]
\centering
\caption*{Table 4. Relative \(L_{2}\) errors of the three complex electric-field components obtained with FNO-Full, FNO-Sparse, and AMNO for test Cases D and E. FNO-Full uses dense reference-field supervision, whereas FNO-Sparse and AMNO use 7.5\% of the reference-field points; only AMNO additionally incorporates the frequency-domain Maxwell-equation residual.}
\begin{tabular}{@{}
  >{\centering\arraybackslash}m{(\linewidth - 8\tabcolsep) * \real{0.1201}}
  >{\centering\arraybackslash}m{(\linewidth - 8\tabcolsep) * \real{0.1886}}
  >{\centering\arraybackslash}m{(\linewidth - 8\tabcolsep) * \real{0.2115}}
  >{\centering\arraybackslash}m{(\linewidth - 8\tabcolsep) * \real{0.2115}}
  >{\centering\arraybackslash}m{(\linewidth - 8\tabcolsep) * \real{0.2116}}@{}}
\toprule\noalign{}
Case & Method & \(E_{R}\) & \(E_{\phi}\) & \(E_{Z}\) \\
\midrule\noalign{}
\multirow{3}{=}{\centering\arraybackslash Case D} & FNO-Full & \(1.12 \times 10^{- 2}\) & \(1.46 \times 10^{- 2}\) & \(1.29 \times 10^{- 2}\) \\
& FNO-Sparse & \(1.93 \times 10^{- 1}\) & \(2.02 \times 10^{- 1}\) & \(2.05 \times 10^{- 1}\) \\
& AMNO & \(2.87 \times 10^{- 2}\) & \(6.84 \times 10^{- 2}\) & \(2.18 \times 10^{- 2}\) \\
\midrule\noalign{}
\multirow{3}{=}{\centering\arraybackslash Case E} & FNO-Full & \(1.05 \times 10^{- 2}\) & \(1.50 \times 10^{- 2}\) & \(1.23 \times 10^{- 2}\) \\
& FNO-Sparse & \(1.99 \times 10^{- 1}\) & \(2.14 \times 10^{- 1}\) & \(2.11 \times 10^{- 1}\) \\
& AMNO & \(2.93 \times 10^{- 2}\) & \(7.02 \times 10^{- 2}\) & \(2.13 \times 10^{- 2}\) \\
\bottomrule\noalign{}
\end{tabular}

\par\vspace{7pt}
\centering
\caption*{Table 5. Single-case solution times of AMNO and COMSOL for test Cases D and E. The AMNO values correspond to post-training forward-inference times, while the COMSOL values are the means of three independent runs.}
\begin{tabular}{@{}
  >{\centering\arraybackslash}m{(\linewidth - 6\tabcolsep) * \real{0.1200}}
  >{\centering\arraybackslash}m{(\linewidth - 6\tabcolsep) * \real{0.3400}}
  >{\centering\arraybackslash}m{(\linewidth - 6\tabcolsep) * \real{0.2700}}
  >{\centering\arraybackslash}m{(\linewidth - 6\tabcolsep) * \real{0.2120}}@{}}
\toprule\noalign{}
Case & AMNO inference time (s) & COMSOL solution time (s) & Speed-up factor \\
\midrule\noalign{}
Case D & 0.2502 & 312 & \(1.25 \times 10^{3}\) \\
Case E & 0.2487 & 327 & \(1.31 \times 10^{3}\) \\
\bottomrule\noalign{}
\end{tabular}
\end{figure*}

\section{Experiments}

In this study, COMSOL solutions of the frequency-domain Maxwell equation with the local anisotropic dielectric model described in section 2.1 served as the reference solutions. The associated finite-element electromagnetic formulation has previously been benchmarked in ICRF antenna--plasma coupling studies against TOPICA, experimental results, and other full-wave solvers {[}19,20{]}. All experiments were conducted on the same \(R\text{--}Z\) computational grid and within the same valid computational domain. A set of cases was constructed over \(X_{H} = 0.01\sim 0.05\), sampling continuous dielectric variation within the low-\(X_{H}\) hydrogen-minority-heating regime considered here rather than a transition between distinct ICRH heating regimes.

The predictive accuracy of AMNO was quantitatively evaluated using the mean squared error (MSE), mean absolute error (MAE), and relative \(L_{2}\) error. All complex-field errors were computed using the complex modulus, thereby accounting for deviations in both the real and imaginary parts. Pointwise spatial errors were normalised by the root-mean-square (RMS) value of the reference field to avoid abnormally inflated relative errors where the local reference field approaches zero. The coefficient of determination (\(R^{2}\)) in the parity plots (reference-versus-prediction scatter plots) was instead calculated from the electric-field magnitude and was used solely to assess agreement in magnitude.

\subsection{Electromagnetic-field solution for a representative case}

To verify the fundamental capability of AMNO to reconstruct the ICRH complex electric-field vector, this section presents a validation experiment for a representative case. The evaluation was performed at three levels: two-dimensional magnitude distributions, radial complex-field line profiles, and full-field errors. Figure 3 compares the COMSOL reference fields, the fields predicted by AMNO, and the corresponding absolute complex-field errors. All three reference electric-field components exhibit pronounced spatial non-uniformity, forming multiscale lobes, local extrema, and alternating oscillatory patterns within the plasma. AMNO reproduces well the principal high-amplitude regions, lobe locations, and overall oscillatory structures of \(E_{R}\) and \(E_{Z}\). The more pronounced errors are concentrated primarily around local peaks and the interfaces between adjacent lobes.

The magnitude of \(E_{\phi}\) is substantially lower than those of the other two components, but it still exhibits clear spatial oscillations. The prediction preserves its principal high-amplitude regions and lobe contours, and no tendency for the weak-amplitude component to degenerate towards a zero field or become excessively smoothed is observed.

The two-dimensional magnitude maps primarily characterise the wave-field envelope and spatial topology, but cannot fully represent sign changes or phase-related information in the complex electric field. To further compare the local oscillatory structures, radial line profiles were extracted at the midplane \(Z = 0\), and the real part, imaginary part, and magnitude of each component are presented separately in figure 4. The curves predicted by AMNO generally follow the corresponding COMSOL reference curves in terms of the principal peak and trough locations, oscillation periods, and variations in the real and imaginary parts. For the dominant-amplitude components \(E_{R}\) and \(E_{Z}\), both sets of curves are in good agreement over most of the radial interval. The deviations occur mainly at rapid transitions between adjacent peaks and troughs and at locations where the field varies sharply, corresponding to the regions of concentrated error observed in figure 3. For the weaker-amplitude component \(E_{\phi}\), the model still recovers the principal oscillatory trend, although its local relative deviations are larger than those of the other two components.

To reduce the influence of a single line profile location on the evaluation, the MSE, MAE, and relative \(L_{2}\) error were further calculated over all valid grid points. The results are listed in table 1. The relative \(L_{2}\) errors are \(2.87 \times 10^{- 2}\) for \(E_{R}\) and \(2.35 \times 10^{- 2}\) for \(E_{Z}\), indicating high full-field reconstruction accuracy for the two principal components in the poloidal plane. The relative \(L_{2}\) error of the toroidal component \(E_{\phi}\) is \(6.96 \times 10^{- 2}\), which is higher than those of the other two components, although its absolute errors remain low. This difference is mainly associated with the smaller reference-field norm of \(E_{\phi}\): when the absolute deviations are similar, a smaller denominator results in a larger relative error. Therefore, weak-amplitude components should be evaluated by considering both relative and absolute errors.

\subsection{Unified operator-based solution across multiple cases}

Building on the preceding single-case validation, we further examine the capability of AMNO to provide a unified representation of different plasma dielectric conditions. A training set comprising multiple discrete modelling cases was constructed over the parameter range \(0.01{\leq X}_{H} \leq 0.05\). All cases used the same computational geometry, radio frequency, toroidal mode number, and spatial grid; only the complex dielectric-tensor field generated from the hydrogen minority fraction corresponding to each case was varied. The following analysis evaluates the model's multi-case solution capability at two levels: (i) wave-field reconstruction for representative cases and cross-case consistency of prediction errors; (ii) local error characteristics and the preservation of spatial scales.

\subsubsection{Multi-case wave-field reconstruction and cross-case consistency}

Cases A, B, and C were selected as representative modelling cases to cover both endpoints and the interior of the parameter range and to enable comparison of the wave-field response differences induced by variations in plasma dielectric conditions using only a limited number of displayed cases. Their hydrogen minority fractions were 0.01, 0.02, and 0.05, corresponding to the lower end, an interior value, and the upper end of the parameter range, respectively.

Among the three electric-field components, \(E_{R}\) has a comparatively large magnitude, and its spatial distribution varies appreciably with the plasma dielectric condition. Figure 5 therefore takes \(\left| E_{R} \right|\) as a representative component and compares the COMSOL reference magnitudes, the AMNO predicted magnitudes, and the corresponding absolute complex-field errors for the three cases. The overall predictive performance for all three complex electric-field components is further evaluated through the subsequent error statistics and spatial-scale analysis.

In figure 5, the reference fields for the three cases differ markedly in peak magnitude, the locations of high-amplitude regions, and lobe topology, reflecting the modulation of radial wave propagation and interference patterns by variations in plasma dielectric conditions. AMNO can generate field distributions characteristic of the corresponding cases in response to different dielectric-tensor inputs, with the principal lobe locations, high-amplitude regions, and overall spatial topology varying accordingly as the plasma dielectric condition changes.

To assess the consistency of AMNO prediction accuracy across the parameter domain, eight cases were randomly sampled from the set of modelling cases and indexed as cases 1--8 in ascending order of hydrogen minority fraction. For each case, the RMS-normalised pointwise complex-field errors of \(E_{R}\), \(E_{\phi}\), and \(E_{Z}\) were calculated at all valid grid points inside the LCFS, and their distributions were compared using box plots, as shown in figure 6. The error medians and interquartile ranges remain at comparable levels across the cases, with no systematic elevation in the error level observed with variations in the hydrogen minority fraction, indicating that AMNO maintains consistent fitting accuracy across the modelling cases. For most boxes, the mean is higher than the median, indicating right-skewed error distributions: most grid points exhibit relatively small errors, whereas larger deviations are concentrated within a limited number of local regions. This statistical pattern is consistent with the concentration of errors around local extrema and lobe interfaces observed in figure 5.

Notably, \(E_{\phi}\) exhibits a higher median error, a broader interquartile range, and larger maximum errors than \(E_{R}\) and \(E_{Z}\). Its higher error is not attributable solely to a very small number of anomalous points, but instead to the larger normalised deviations exhibited by the weak-amplitude component over a broader spatial region.

\subsubsection{Local error characteristics and spatial-scale representation}

The absolute errors in figure 5 exhibit pronounced spatial non-uniformity, with larger deviations occurring primarily near local extrema, in regions of rapid field variation, and in the boundary regions between adjacent fine-scale lobes. To quantify the relationship between local field variation and prediction error, modelling case A is considered, and the statistical association between the local gradient of the reference-field magnitude and the pointwise complex-field error of AMNO is examined.

For an electric-field component \(E_{j}\), where \(j \in \left\{ R,\phi,Z \right\}\), the local gradient of the reference-field magnitude is defined as

\begin{equation*}
\begin{array}{r}
g_{j}(R,Z) = \left| \nabla\left| E_{j}^{COMSOL}(R,Z) \right| \right|,
\end{array}
\tag{19}
\end{equation*}

and is normalised by its 95th percentile:

\begin{equation*}
\begin{array}{r}
{\widetilde{g}}_{j} = \frac{g_{j}}{P_{95}\left( g_{j} \right)}.
\end{array}
\tag{20}
\end{equation*}

Figure 7 uses hexagonal binning to show the joint gradient--error distribution over the valid grid points. It also presents the median and interquartile range of the errors in each of 10 gradient intervals containing equal numbers of samples, together with the Spearman rank correlation coefficient. The reference-field gradient and the pointwise prediction error exhibit a positive statistical association for all three electric-field components, with Spearman rank correlation coefficients of 0.424, 0.435, and 0.391, respectively. As the normalised reference-field gradient increases, the median error of each component generally increases, while the interquartile range broadens markedly in the high-gradient intervals. These results show that regions of rapid field variation not only have larger typical prediction errors but also greater error dispersion across spatial locations.

Compared with \(E_{R}\) and \(E_{Z}\), \(E_{\phi}\) exhibits a higher error median and a broader interquartile range, indicating more pronounced local prediction difficulties for the weak-amplitude toroidal component in high-gradient regions. This is consistent with the preceding analysis of the error metrics for \(E_{\phi}\).

It should be noted that a certain degree of vertical error dispersion remains within the same gradient range, indicating that the reference-field gradient is not the only factor determining prediction accuracy. The local field magnitude, spatial-wavenumber content, and dynamic ranges of the different components may also jointly affect the prediction error.

To further examine the capability of AMNO to represent different spatial scales, the three complex electric-field components are transformed from the spatial domain to the two-dimensional wavenumber domain. Field values outside the valid region are set to zero, and the mean is removed and a two-dimensional window function is applied before the Fourier transform to reduce spectral leakage caused by the finite computational domain. The spectral magnitude is normalised by the maximum spectral magnitude of the corresponding COMSOL component and expressed in dB.

In figure 8, the dominant spectral energy of all three components is concentrated in the low-to-intermediate-wavenumber region. AMNO closely recovers the central locations, principal directions of extension, and dominant energy bands of the reference spectra, demonstrating that the method not only reconstructs the macroscopic topology of the spatial fields but also preserves their principal spatial-scale composition. The predicted spectra of \(E_{R}\) and \(E_{Z}\) are in good agreement with their reference spectra, whereas \(E_{\phi}\) shows more pronounced differences in spectral bandwidth and magnitude near \(k_{Z} \approx 0\), consistent with its higher relative field error.

\subsection{In-range interpolation at held-out \(\boldsymbol{X}_{\boldsymbol{H}}\) values}

We next evaluate the in-range interpolation accuracy of AMNO for held-out dielectric conditions belonging to the same \(X_{H}\)-generated family as the modelling cases. Cases D and E were excluded from training and correspond to \(X_{H} = 0.025\) and \(0.038\), respectively, both of which lie within the modelled interval \(0.01{\leq X}_{H} \leq 0.05\). These cases therefore assess interpolation to held-out \(X_{H}\) values under otherwise unchanged physical and computational settings, rather than parameter extrapolation, cross-regime generalisation, or transfer to dielectric-tensor fields outside the modelled family.

To avoid repetition, figures 9 and 10 present the prediction results only for Case D, while the overall accuracy for both test cases is subsequently compared in figure 11 and tables 2 and 3. Figure 9 presents the COMSOL reference magnitudes, the AMNO predicted magnitudes, and the corresponding absolute complex-field errors for the three electric-field components in test case D. For this unseen plasma dielectric condition within the covered parameter range, AMNO still recovers the principal lobe locations and overall spatial topology of all three electric-field components. The more pronounced errors occur primarily in local high-gradient regions and at the boundaries between fine-scale lobes, and no large-scale oscillatory structures that are absent from the reference solution are introduced. This result demonstrates that AMNO has good predictive capability for case D, an unseen interpolative test case.

To further evaluate the ability of the operator to jointly recover the magnitudes and relative phase relationship of the poloidal complex electric-field components, two polarisation-combination fields are constructed from the predicted components:

\begin{equation*}
\begin{array}{r}
E_{+} = E_{R} + iE_{Z},\ \ E_{-} = E_{R} - iE_{Z}.
\end{array}
\tag{21}
\end{equation*}

Both \(E_{+}\) and \(E_{-}\) are obtained by post-processing the \(E_{R}\) and \(E_{Z}\) components predicted by the network. Because they are formed through the complex superposition of the two poloidal complex-field components, their magnitudes depend jointly on the magnitudes of \(E_{R}\) and \(E_{Z}\) and on their relative phase. Therefore, a further comparison of the polarisation-combination fields reconstructed from the predicted complex fields is conducted using Case D as a representative case. Figure 10 presents the COMSOL reference and AMNO reconstructed magnitudes of \(E_{+}\) and \(E_{-}\), together with the corresponding absolute errors.

The AMNO reconstructions preserve the principal high-magnitude regions, lobe locations, and overall oscillatory topology of both polarisation-combination fields. The more pronounced absolute errors remain concentrated primarily near local peaks, in regions of rapid transitions between lobes, and in regions adjacent to the LCFS. These characteristics are consistent with those observed for the three electric-field components in figure 9 and in the gradient--error statistics in figure 7.

It should be emphasised that these results demonstrate the ability of the model to recover the spatial structures of the polarisation-combination fields, but cannot be used alone to infer ion power absorption or heating efficiency. The relevant physical quantities still require further calculation in conjunction with power deposition or wave--particle interaction models.

Figures 9 and 10 present the spatial reconstruction results for Case D from the perspectives of the cylindrical-coordinate components and the polarisation-combination fields, respectively. However, the spatial maps mainly reveal the local lobe topology and the locations of the errors, making it difficult to quantitatively assess the magnitude correlation over all valid grid points or to directly compare the overall prediction consistency of the two unseen cases. Figure 11 therefore presents COMSOL--AMNO parity plots for Case D and Case E. Each point corresponds to one grid point, with the COMSOL electric-field magnitude on the horizontal axis and the AMNO electric-field magnitude on the vertical axis. The dashed line \(y = x\) represents the ideal prediction. Most of the points form narrow bands along the reference line, indicating that the predictions preserve the magnitude ordering and principal dynamic ranges of the reference fields. For both cases, the \(R^{2}\) values of all three components exceed 0.99. The scatter bands for \(E_{R}\) and \(E_{Z}\) are reasonably narrow, indicating good agreement for the two principal components at both low and high field magnitudes. The scatter for \(E_{\phi}\) is more dispersed, consistent with its lower overall field magnitude and higher normalised error.

Table 2 further reports the complex-field MSE, MAE, and relative \(L_{2}\) errors for the two test cases over all valid grid points within the LCFS. On this basis, to determine where the test-case errors lie relative to the error distribution of the modelling cases, table 3 compares the relative \(L_{2}\) errors of the test cases with the overall statistical levels of the relative \(L_{2}\) errors for the modelling cases, expressed in terms of their means and standard deviations.

Taken together, tables 2 and 3 show that Case D and Case E have similar error levels for all three electric-field components. Their errors are all close to the means of the modelling cases, and their deviations from these means do not exceed one inter-case standard deviation. The results show that the prediction errors for the test cases exhibit no degradation, further confirming the stable and reliable predictive capability of AMNO within the current parameter range.

\subsection{Benefits of the physics constraint and computational efficiency under limited reference-field supervision}

The preceding experiments have demonstrated the capability of AMNO to predict complex electric fields for unseen interpolation test cases. However, they do not distinguish whether its predictive accuracy arises from the Fourier neural operator backbone or from the explicitly imposed frequency-domain Maxwell-equation constraint. To disentangle the contributions of reference-field supervision density and the physics constraint, three methods---FNO-Full, FNO-Sparse, and AMNO---are compared on the same test cases in this section.

All three methods use the same FNO backbone. FNO-Full is a densely supervised model trained using all available reference-field points and does not include the frequency-domain Maxwell-equation residual. Starting from this configuration, FNO-Sparse is obtained by reducing the number of reference-field points to 7.5\% of that used by FNO-Full. It is likewise optimised solely using the reference-field consistency loss and does not include the frequency-domain Maxwell-equation residual. AMNO uses the same limited set of discrete reference-field points as FNO-Sparse but additionally incorporates the frequency-domain Maxwell-equation residual alongside the reference-field consistency constraint. The relative \(L_{2}\) errors of the three complex electric-field components obtained using the three methods for test cases D and E are reported in table 4. With only the reference-field consistency constraint imposed, the predictive accuracy of FNO is relatively sensitive to the density of the reference-field points. When the number of reference-field points is reduced to 7.5\% of the dense configuration, the errors of FNO-Sparse for all three electric-field components in both test cases increase to approximately 20\%. This indicates that a limited set of discrete reference-field points is insufficient to fully recover the lobe distributions, local oscillations, and interrelationships among the three electric-field components over the entire poloidal cross-section.

Under the same limited-reference-field supervision, AMNO makes more efficient use of the available reference-field information. Relative to FNO-Sparse, AMNO reduces the six relative \(L_{2}\) errors by 66.1\%--89.9\%, with consistent improvements observed for both unseen cases. These results indicate that the improvement in accuracy arises from the ability of the frequency-domain Maxwell-equation constraint to supplement the limited reference-field information and reduce the dependence of operator learning on dense reference fields, rather than from an increase in model capacity or in the number of labelled reference-field points.

It should be noted that FNO-Full achieves the lowest prediction errors for both unseen cases, indicating that the explicit physics constraint cannot fully replace the solution information contained in densely sampled reference fields. Nevertheless, using only 7.5\% of the reference-field points, AMNO reduces the errors of all three components to the same order of magnitude as those of the densely supervised model. Taking the error gap between FNO-Sparse and FNO-Full as the baseline, AMNO closes approximately 71.3\%--95.5\% of this accuracy gap. Thus, the advantage of AMNO does not lie in outperforming a purely data-driven model under full supervision, but in substantially reducing the dependence of operator learning on dense field supervision through the governing-equation constraint when reference-field information is limited.

Building on the accuracy comparison, the computational efficiency of AMNO for unseen plasma dielectric conditions is further examined. All timings were obtained on the same computational workstation. COMSOL was run independently three times for each case, and the mean solution time was recorded. The AMNO timing corresponds to the forward inference time for a single case after offline training. The results are listed in table 5. Under the present test conditions, AMNO, while substantially reducing its dependence on dense reference-field supervision, shortens the computation time for the electric-field response of unseen dielectric cases from a timescale of minutes to a subsecond timescale and has nearly identical inference times for the two cases. Taken together, tables 4 and 5 show that AMNO is not simply a field-value regression model that relies on densely sampled full-wave solutions. Instead, it uses the frequency-domain Maxwell equation to convert governing-equation information into a domain-wide constraint during operator learning and simultaneously imposes the dielectric tensor, spatial derivatives, coupling among the three electric-field components, and wave-propagation relations on the complete predicted field. Consequently, what the network learns is no longer limited to numerical interpolation between reference points but is constrained to the feasible solution space satisfying the governing equation. AMNO therefore provides a parametric solution approach for ICRH parameter sweeps and multi-case response analysis that combines physical constraints, data efficiency, and rapid online solution capability.

\FloatBarrier
\section{Conclusion}

Accurate prediction of the three-component complex electric field distribution for ICRH in complex magnetised plasmas is essential for analysing wave propagation, calculating power deposition, and optimising heating scenarios. To tackle the repeated full-wave numerical solutions required in multi-case calculations, this work proposes AMNO, which provides shared prediction of the three-component complex electric field across the \(X_{H}\)-generated dielectric conditions considered here while retaining the complex anisotropic dielectric tensor and multiscale wave characteristics.

The results of this study demonstrate that the proposed model accurately reconstructs the principal lobe structures, local extrema, and spatial oscillations of the three-component complex electric field, while maintaining close agreement with the reference solutions in the real and imaginary parts, field magnitudes, and principal spatial-spectral characteristics. The full-field relative \(L_{2}\) errors of the radial and vertical components remain within approximately 2.13\%--2.93\%, indicating high field-reconstruction accuracy. Although the toroidal component exhibits a higher relative error, its absolute errors remain at a relatively low level. The overall errors for the test cases are comparable to those for the modelling cases, and the coefficients of determination for the magnitudes of all three components exceed 0.99. These results indicate that the learned operator can stably characterise the mapping between variations in plasma dielectric conditions and the corresponding complex electric-field responses within the current parameter range.

The baseline comparison under sparse reference-field conditions further confirms the practical contribution of the frequency-domain Maxwell-equation constraint. When the number of reference-field points is reduced to 7.5\% of the dense configuration, AMNO reduces the six relative \(L_{2}\) errors by 66.1\%--89.9\% relative to FNO-Sparse and closes 71.3\%--95.5\% of the accuracy gap between FNO-Sparse and FNO-Full. Although densely sampled reference fields contain solution information that cannot be fully replaced, these results show that the frequency-domain Maxwell-equation residual transforms the dielectric tensor, spatial derivatives, and coupling among the three electric-field components into a domain-wide constraint acting on the complete predicted field. Operator learning is therefore no longer limited to numerical fitting between a finite number of reference points, substantially reducing its dependence on dense full-wave solutions. Following the completion of the one-off offline training and model deployment, the inference time for a single case is approximately 0.25 s, reducing the computation time from the minute scale required for COMSOL full-wave solutions to the subsecond scale under the present test conditions. These results demonstrate the combined advantages of the proposed method in terms of efficient reference-field utilisation and online solution efficiency. The proposed method therefore provides a rapid parametric surrogate for in-range \(X_{H}\) sweeps and cross-case response analysis within the modelled EAST configuration, while integrating physical constraints and limited reference-field supervision.

This work extends physics-informed operator learning to the parametric modelling of ICRH three-component complex electric fields in magnetised plasmas and establishes a physics-constrained surrogate for the one-parameter family of dielectric-tensor fields generated by varying \(X_{H}\) within the modelled interval. To the best of our knowledge, such a physics-informed neural-operator formulation has not previously been reported for ICRH full-wave modelling. This framework thereby provides a new route for rapid surrogate modelling of full-wave ICRH responses across multiple cases. On this basis, the proposed method can convert repeated case-by-case full-wave solutions into a reusable capability for rapid complex-field inference, providing an efficient computational foundation for in-range \(X_{H}\) sweeps and cross-case comparisons within the modelled EAST configuration. It should be noted that the present study is restricted to a fixed radio frequency, toroidal mode, computational geometry, and antenna excitation, and that all unseen test cases considered lie within both the modelled \(X_{H}\) interval and the same low-\(X_{H}\) hydrogen-minority-heating regime. Moreover, although AMNO directly takes \(\boldsymbol{\varepsilon}_{r}(R,Z)\) as input, all dielectric-tensor fields considered here are generated from the single parameter \(X_{H}\) and therefore constitute a low-dimensional one-parameter family. Accordingly, the present results support only in-range interpolation within this heating regime and should not be interpreted as evidence of parameter extrapolation, cross-regime generalisation, or generalisation to arbitrary dielectric-tensor fields. Future work will expand both the number of cases and the dimensionality of the training distribution by independently varying plasma composition, density and temperature profiles, equilibrium, radio frequency, toroidal mode number, geometry, and antenna excitation, and will employ explicit out-of-range and out-of-family test sets to evaluate parameter extrapolation and transfer across more diverse dielectric-field distributions.

\FloatBarrier

\section*{Acknowledgements}

This work was supported by the Science and Technology on Reactor System Design Technology Laboratory (Grant No.~LRSDT12023108) and the Research Start-up Fund Project of Chongqing University of Posts and Telecommunications under Grant Nos.~A2020-217 and A2020-216.

\section*{Data availability}

The data that support the findings of this study are available from the corresponding author upon reasonable request.

\section*{References}
\balance

\begingroup
\small
\begin{enumerate}
\def\labelenumi{[\arabic{enumi}]}
\item
  Wilson J.R. and Bonoli P.T. 2015 Progress on ion cyclotron range of frequencies heating physics and technology in support of the International Tokamak Experimental Reactor \emph{Phys. Plasmas} \textbf{22} 021801
\item
  Zhang X.J. et al 2012 Initial results on plasma heating experiments in the ion cyclotron range of frequencies on EAST \emph{Nucl. Fusion} \textbf{52} 032002
\item
  Van Eester D. et al 2019 Ion cyclotron resonance heating scenarios for DEMO \emph{Nucl. Fusion} \textbf{59} 106051
\item
  Start D.F.H. et al 1999 Bulk ion heating with ICRH in JET DT plasmas \emph{Nucl. Fusion} \textbf{39} 321--336
\item
  Lerche E. et al 2016 Optimization of ICRH for core impurity control in JET-ILW \emph{Nucl. Fusion} \textbf{56} 036022
\item
  Jacquet P. et al 2024 ICRH operations during the JET tritium and DTE2 campaigns \emph{Nucl. Fusion} \textbf{64} 066039
\item
  Zhang J.H. et al 2017 Experimental analysis of the ICRF waves coupling in EAST \emph{Nucl. Fusion} \textbf{57} 066030
\item
  Zhang X.J. et al 2022 First experimental results with new ICRF antenna in EAST \emph{Nucl. Fusion} \textbf{62} 086038
\item
  Zhang J.H. et al 2023 Influences of plasma density perturbations on ion cyclotron resonance heating \emph{Nucl. Fusion} \textbf{63} 046012
\item
  Kazakov Y.O. et al 2012 Study of ICRH scenarios for thermal ion heating in JET D--T plasmas \emph{Nucl. Fusion} \textbf{52} 094012
\item
  Kazakov Y.O. et al 2015 On resonant ICRF absorption in three-ion component plasmas: a new promising tool for fast ion generation \emph{Nucl. Fusion} \textbf{55} 032001
\item
  Kazakov Y.O. et al 2013 Effect of impurities on the transition between minority ion and mode conversion ICRH heating in (3He)--H tokamak plasmas \emph{Nucl. Fusion} \textbf{53} 053014
\item
  Mantsinen M.J. et al 2004 Localized bulk electron heating with ICRF mode conversion in the JET tokamak \emph{Nucl. Fusion} \textbf{44} 33--46
\item
  Brambilla M. 1999 Numerical simulation of ion cyclotron waves in tokamak plasmas \emph{Plasma Phys. Control. Fusion} \textbf{41} 1--34
\item
  Jaeger E.F. et al 2001 All-orders spectral calculation of radio-frequency heating in two-dimensional toroidal plasmas \emph{Phys. Plasmas} \textbf{8} 1573--1583
\item
  Jaeger E.F. et al 2002 Advances in full-wave modeling of radio frequency heated, multidimensional plasmas \emph{Phys. Plasmas} \textbf{9} 1873--1881
\item
  Brambilla M. and Bilato R. 2006 Simulation of ion cyclotron heating of tokamak plasmas using coupled Maxwell and quasilinear-Fokker--Planck solvers \emph{Nucl. Fusion} \textbf{46} S594--S614
\item
  Jucker M. et al 2011 Integrated modeling for ion cyclotron resonant heating in toroidal systems \emph{Comput. Phys. Commun.} \textbf{182} 912--925
\item
  Tierens W. et al 2019 Validation of the ICRF antenna coupling code RAPLICASOL against TOPICA and experiments \emph{Nucl. Fusion} \textbf{59} 046001
\item
  Budny R.V. et al 2012 Benchmarking ICRF full-wave solvers for ITER \emph{Nucl. Fusion} \textbf{52} 023023
\item
  Usoltceva M. et al 2019 Simulation of the ion cyclotron range of frequencies slow wave and the lower hybrid resonance in 3D in RAPLICASOL \emph{Plasma Phys. Control. Fusion} \textbf{61} 115011
\item
  Yin L. et al 2024 A full wave solver integrated with a Fokker--Planck code for optimizing ion heating with ICRF waves for the ITER deuterium--tritium plasma \emph{Nucl. Fusion} \textbf{64} 076020
\item
  Bertelli N. et al 2017 Full-wave simulations of ICRF heating regimes in toroidal plasma with non-Maxwellian distribution functions \emph{Nucl. Fusion} \textbf{57} 056035
\item
  Shiraiwa S. et al 2017 HIS-TORIC: extending core ICRF wave simulation to include realistic SOL plasmas \emph{Nucl. Fusion} \textbf{57} 086048
\item
  Perkins R.J. et al 2019 Resolving interactions between ion-cyclotron range of frequencies heating and the scrape-off layer plasma in EAST using divertor probes \emph{Plasma Phys. Control. Fusion} \textbf{61} 045011
\item
  Raissi M. et al 2019 Physics-informed neural networks: A deep learning framework for solving forward and inverse problems involving nonlinear partial differential equations \emph{J. Comput. Phys.} \textbf{378} 686--707
\item
  Karniadakis G.E. et al 2021 Physics-informed machine learning \emph{Nat. Rev. Phys.} \textbf{3} 422--440
\item
  Lu L. et al 2021 Learning nonlinear operators via DeepONet based on the universal approximation theorem of operators \emph{Nat. Mach. Intell.} \textbf{3} 218--229
\item
  Wang S. et al 2021 Learning the solution operator of parametric partial differential equations with physics-informed DeepONets \emph{Sci. Adv.} \textbf{7} eabi8605
\item
  Li Z. et al 2021 Fourier neural operator for parametric partial differential equations \emph{Int. Conf. on Learning Representations}
\item
  Kovachki N. et al 2023 Neural operator: Learning maps between function spaces with applications to PDEs \emph{J. Mach. Learn. Res.} \textbf{24(89)} 1--97
\item
  Li Z. et al 2024 Physics-informed neural operator for learning partial differential equations \emph{ACM/IMS J. Data Sci.} \textbf{1(3)} 1--27
\item
  Azizzadenesheli K. et al 2024 Neural operators for accelerating scientific simulations and design \emph{Nat. Rev. Phys.} \textbf{6} 320--328
\item
  Zhu M. et al 2023 Reliable extrapolation of deep neural operators informed by physics or sparse observations \emph{Comput. Methods Appl. Mech. Eng.} \textbf{412} 116064
\item
  Pestourie R. et al 2023 Physics-enhanced deep surrogates for partial differential equations \emph{Nat. Mach. Intell.} \textbf{5} 1458--1465
\item
  Lim J. and Psaltis D. 2022 MaxwellNet: Physics-driven deep neural network training based on Maxwell's equations \emph{APL Photonics} \textbf{7} 011301
\item
  Peng Z. et al 2022 Rapid surrogate modeling of electromagnetic data in frequency domain using neural operator \emph{IEEE Trans. Geosci. Remote Sens.} \textbf{60} 2007912
\item
  Augenstein Y. et al 2023 Neural operator-based surrogate solver for free-form electromagnetic inverse design \emph{ACS Photonics} \textbf{10} 1547--1557
\item
  Li Z. et al 2023 Fourier neural operator with learned deformations for PDEs on general geometries \emph{J. Mach. Learn. Res.} \textbf{24(388)} 1--26
\item
  Bonotto M. et al 2024 Reconstruction of plasma equilibrium and separatrix using convolutional physics-informed neural operator \emph{Fusion Eng. Des.} \textbf{200} 114193
\item
  Sánchez-Villar Á. et al 2024 Real-time capable modeling of ICRF heating on NSTX and WEST via machine learning approaches \emph{Nucl. Fusion} \textbf{64} 096039
\end{enumerate}
\endgroup

\end{document}